\documentclass{article}
\renewenvironment{abstract}{\section*{Abstract}\small}{}
\newtheorem{definition}{Definition}

\newtheorem{lemma}[definition]{Lemma}
\newtheorem{notation}[definition]{Notation}

\newtheorem{proposition}[definition]{Proposition}

\newtheorem{remark}[definition]{Remark}

\makeatletter
\renewcommand{\@begintheorem}[2]{ % not in italics
\trivlist\item[\hskip\labelsep{\bf #1\ #2}]}
\renewcommand{\@opargbegintheorem}[3]{\trivlist
\item[\hskip \labelsep{\bf #1\ #2\ (#3)}]}
\makeatother
\newtheorem{proof}{Proof}
\usepackage{multirow}
\usepackage{amssymb}
\newcommand{\qed}{\nobreak \ifvmode \relax \else
\ifdim\lastskip<1.5em \hskip-\lastskip
\hskip1.5em plus0em minus0.5em \fi \nobreak
\vrule height0.75em width0.5em depth0.25em\fi}
      
\newcommand{\M}[1]{M_{#1}}
\newcommand{\MM}[2]{M_{#1, #2}}
\newcommand{\MMM}[3]{M_{#1, #2, #3}}
\newcommand{\monH}[1]{H(#1)}
\newcommand{\monHi}[2]{H_{#1}(#2)}
\newcommand{\T}[1]{T(#1)}
\newcommand{\TT}[2]{T(#1, #2)}
\newcommand{\Td}[1]{T_d(#1)}
\newcommand{\C}[2]{{#1}(#2)}
\newcommand{\CC}[3]{{#1}(#2, #3)}
\newcommand{\CCC}[4]{{#1}(#2, #3, #4)}
\newcommand{\CCCC}[5]{{#1}(#2, #3, #4, #5)}
\newcommand{\nup}[1]{\mu_{#1}}
\newcommand{\nupp}[2]{\mu_{#1}(#2)}

\author{Jonathan Ben-Naim
\hfill \\
LIF, CNRS\hfill \\
CMI 39, rue Joliot-Curie\hfill \\
F-13453 Marseille Cedex 13, France\hfill \\
jbennaim@lif.univ-mrs.fr}
\title{Pivotal and Pivotal-discriminative Consequence Relations
\footnote{
This is an updated version of the paper of the same title published in {\it The Journal of Logic and Computation}.
This version just contains a better presentation (so the numbering of definitions and propositions is different).}}
\date{}

\begin{document}
\maketitle

\begin{abstract}
In the present paper, we investigate
consequence relations that are both paraconsistent and plausible
(but still monotonic). More precisely, we put the focus on pivotal consequence relations, i.e. those relations that can be defined by a pivot (in the style of e.g. D.~Makinson).
A pivot is a fixed subset of valuations
which are considered to be the important ones in the absolute sense.
We worked with a general notion of valuation that covers e.g. the classical valuations
as well as certain kinds of many-valued valuations.
In the many-valued cases, pivotal consequence relations are paraconsistant (in addition to be plausible),
i.e. they are capable of drawing reasonable conclusions which contain contradictions.
We will provide in our general framework syntactic characterizations of several families
of pivotal relations. In addition, we will provide, again in our general framework,
characterizations of several families of pivotal-discriminative consequence relations.
The latter are defined exactly as the plain version, but contradictory conclusions are rejected.
We will also answer negatively a representation problem that was left open by Makinson.
Finally, we will put in evidence a connexion with $X$-logics from Forget, Risch, and Siegel.
The motivations and the framework of the present paper are very close to those of
a previous paper of the author which is about preferential consequence relations.
\end{abstract}

%\newpage
%\tableofcontents
\newpage

%%%%%%%%%%%%%%%%%%%%%%%%%%%%%%%%%%%
\section{Introduction} \label{PIVintro}

One of the main motivations of this paper is to combine some tools used in Paraconsistent Reasoning on the one hand
and Plausible Reasoning on the other hand to deal with both incomplete and inconsistent information.
Actually, the motivations of the present paper are very close to those of a previous paper of the author
which is about preferential consequence relations \cite{Bennaim1}.

Many-valued consequence relations have been developed with the aim of dealing with inconsistent information.
These relations are defined in frameworks where valuations
can assign more than two different truth values to formulas.
In fact, they tolerate contradictions within the conclusions, but reject
the principle of explosion according to which a single contradiction entails the deduction
of every formula.

Independently, plausible (generally non-monotonic) consequence relations have been developed with the aim of dealing with incomplete information. {\it Choice functions} are central tools to define plausible relations.
Indeed, suppose we have at our disposal a function $\mu$, called a choice function, which chooses
in any set of valuations $V$, those elements which are preferred,
not necessarily in the absolute sense, but when the valuations in $V$ are the only ones
under consideration. Then, we can define a plausible consequence relation in the following natural way:
a formula $\alpha$ follows from a set of formulas $\Gamma$ iff
every model for $\Gamma$ chosen by $\mu$ is a model for $\alpha$.

In the present paper, we put the focus on a particular family of choice functions.
Let us present it.
Suppose some valuations are considered to be the important ones in the absolute sense
and collect them in a set $\cal I$, called a pivot. This defines naturally a choice function.
Indeed, simply choose in any set of valuations, those elements that belong to $\cal I$.
Those choice functions which can be defined in this manner constitute the aforementioned family.
The consequence relations defined by this family are called {\it pivotal consequence relations}.
Their importance has been put in evidence by D. Makinson
in \cite{Makinson1, Makinson2} where it is shown that they constitute an
easy conceptual passage between classical and plausible non-monotonic relations.
Indeed, they are perfectly monotonic but already display some of the distinctive
features (i.e. the choice functions) of plausible non-monotonic relations.

For a long time, research efforts on paraconsistent relations and plausible relations
were separated. However, in many applications, the information is both incomplete and
inconsistent. For instance, the semantic web or big databases inevitably contain inconsistencies.
This can be due to human or material imperfections as well as contradictory sources of information.
On the other hand, neither the web nor big databases can contain ``all'' information.
Indeed, there are rules of which the exceptions cannot be enumerated.
Also, some information might be left voluntarily vague or in concise form.
Consequently, consequence relations that are both paraconsistent and plausible are useful
to reason in such applications.

Such relations first appear in e.g. \cite{Priest1, Batens1, KiferLozinskii1, ArieliAvron4, MarquisKonieczny1}.
The idea begins by taking a many-valued framework to get paraconsistency.
Then, only those models that are most preferred according to
some particular binary preference relation on valuations (in the style of \cite{Shoham1, Shoham2})
are relevant for making inference,
which provides plausibility (and in fact also non-monotonicity).
In \cite{AvronLev2, AvronLev3},
A. Avron and I. Lev generalized the study to families of binary preference relations which compare two valuations
using, for each of them, this part of a certain set of formulas it satisfies.
The present paper follows this line of research by combining many-valued frameworks
and choice functions.

More explicitly, we will investigate pivotal consequence relations in a general framework.
According to the different assumptions which will be made about the latter,
it will cover various kinds of frameworks, including e.g. the classical propositional one
as well as some many-valued ones.
Moreover, in the many-valued frameworks, pivotal relations lead to non-trivial
conclusions is spite of the presence of contradictions and are thus useful to deal with
both incomplete and inconsistent information. However, they will not satisfy
the Disjunctive Syllogism (from $\alpha$ and $\neg\alpha \vee \beta$ we can conclude $\beta$),
whilst they satisfy it in classical frameworks.

In addition, we will investigate {\it pivotal-discriminative consequence relations}.
They are defined exactly as the plain version, but any conclusion such that its negation is also a conclusion is rejected.
In the classical framework, they do not bring something really new.
Indeed, instead of concluding everything in the face of inconsistent information,
we will simply conclude nothing.
On the other hand, in the many-valued frameworks, where
the conclusions are rational even from inconsistent information,
the discriminative version will reject the contradictions among them, rendering them
all the more rational.

As a first contribution, we will characterize, in our general framework, several families of pivotal(-discriminative) consequence relations. To do so, we will use techniques very similar to those of a previous paper of the author \cite{Bennaim1}.
The latter is about another family of choice functions.
Let us present it.
Suppose we are given a binary preference relation $\prec$ on states labelled by valuations
(in the style of e.g. \cite{KrausLehmannMagidor1, LehmannMagidor1, Schlechta5}).
This defines naturally a choice function.
Indeed, choose in any set of valuations $V$, each element which labels a state which is $\prec$-preferred
among those states which are labelled by the elements of $V$.
Those choice functions which can be defined in this manner constitute the aforementioned family.
The consequence relations defined by this family of choice functions are called
{\it preferential(-discriminative) consequence relations}.
In fact, the present paper provides an example of how the techniques developed in \cite{Bennaim1}
(especially, in the discriminative case) can be adapted to new families of choice functions.
Note that, in the non-discriminative case, the techniques of \cite{Bennaim1} are themselves
strongly inspired by the work of K. Schlechta \cite{Schlechta5}.

In many cases, our characterizations will be purely syntactic.
This has a lot of advantages, let us quote some important ones.
Take a set of syntactic conditions that characterizes a family of pivotal consequence relations.
This gives a syntactic point of view on this family defined semantically,
which enables us to compare it to conditions known on the ``market'',
and thus to other consequence relations.
This can also give rise to questions like:
if we modified the conditions in such and such a natural-looking way,
what would happen on the semantic side?
More generally, this can open the door to questions that would not easily come
to mind otherwise or to techniques of proof that could not have been employed in the
semantic approach.

Some characterizations of pivotal consequence relations, valid in classical frameworks,
can be found in the literature, e.g. \cite{Rott1, Makinson1, Makinson2}.
But, to the author knowledge, the present paper contains the first systematic work of
characterization for them in non-classical frameworks.
Similarly, it seems that the author is the first to investigate pivotal-discriminative consequence relations.

As a second contribution, we will answer negatively a representation problem
that was left open by Makinson, namely,
in an infinite classical framework, there does not exist a ``normal'' characterization for the family of all pivotal
consequence relations. Approximatively, a characterization is called normal
iff it contains only conditions universally quantified and of limited size.
This constitutes the more innovative part of the paper.
A last contribution is that a certain family of pivotal consequence relations will be shown to
be precisely a certain family of $X$-logics, which were introduced by Forget, Risch, and Siegel \cite{ForgetRischSiegel1}.

The rest of the paper is organized as follows.
In Section~\ref{PIVframework}, we introduce our general framework
and the different assumptions which will sometimes be made about it.
We will see that it covers in particular the many-valued frameworks of the well-known paraconsistent logics
$\cal FOUR$ and $J_3$.
In Section~\ref{PIVchoicefun}, we present choice functions and some of their well-known properties.
We will see which properties characterize those choice functions that can be defined by a pivot.
In Section~\ref{PIVintroprefCR}, we define pivotal(-discriminative) consequence
relations and give examples of them in both the classical and the many-valued frameworks.
In Section~\ref{PIVcacarac}, we provide our characterizations.
In Section~\ref{PIVsectionNoCharac}, we answer negatively the problem that was left open by Makinson.
In Section~\ref{PIVsectioncodefisclosed}, we put in evidence a connexion
with $X$-logics.
Finally, we conclude in Section~\ref{PIVconclu}.
  
%%%%%%%%%%%%%%%%%%%%%%%%%%%%%%%%%%%
\section{Background} \label{PIVprelim}

%%%%%%%%%%%%%%%%%%%%%%%%
\subsection{Semantic structures} \label{PIVframework}

%%%%%%%%%%%%%%%%%%%%%%%%
\subsubsection{Definitions and properties}

The framework is exactly the one presented in \cite{Bennaim1}.
We will work with general formulas, valuations, and satisfaction.
A similar approach has been taken in two well-known papers \cite{Makinson2, Lehmann2}.

\begin{definition}
We say that $\cal S$ is a {\it semantic structure} iff
${\cal S} = \langle {\cal F}, {\cal V}, \models \rangle$ where
$\cal F$ is a set,
$\cal V$ is a set,
and $\models$ is a relation on ${\cal V} \times {\cal F}$.
\end{definition}
Intuitively, $\cal F$ is a set of formulas,
$\cal V$ a set of valuations for these formulas, and
$\models$ a satisfaction relation for these objects (i.e. $v \models \alpha$ means the formula $\alpha$ is satisfied in the valuation $v$,
i.e. $v$ is a model for $\alpha$).

\begin{notation}
Let $\langle {\cal F}, {\cal V}, \models \rangle$ be a semantic structure,
$\Gamma \subseteq {\cal F}$, and $V \subseteq {\cal V}$. Then,
\hfill \\
$\M{\Gamma} := \lbrace v \in {\cal V} : \forall \: \alpha \in \Gamma$, $v \models \alpha \rbrace$,
\hfill \\
$\T{V} := \lbrace \alpha \in {\cal F} : V \subseteq \M{\alpha} \rbrace$,
\hfill \\
${\bf D} := \lbrace V \subseteq {\cal V} : \exists \: \Gamma \subseteq {\cal F}, \M{\Gamma} = V \rbrace$.
\hfill \\
Suppose $\cal L$ is a language, $\neg$ a unary connective of $\cal L$, and 
$\cal F$ the set of all wffs of $\cal L$. Then,
\hfill \\
$\Td{V} := \lbrace \alpha \in {\cal F} : V \subseteq \M{\alpha}$ and $V \not \subseteq \M{\neg\alpha} \rbrace$,
\hfill \\
${\bf C} := \lbrace V \subseteq {\cal V} : \forall \: \alpha \in {\cal F}$, $V \not\subseteq \M{\alpha}$ or
$V \not\subseteq \M{\neg\alpha} \rbrace$.
\end{notation}
Intuitively, $\M{\Gamma}$ is the set of all models for $\Gamma$
and $\T{V}$ the set of all formulas satisfied in $V$.
Roughly speaking, $\Td{V}$ is this part of $\T{V}$ that is not contradictory.
$\bf D$ is the set of all those sets of valuations that are definable by a set of formulas
and $\bf C$ the set of all those sets of valuations that do not satisfy both a formula and its negation.
As usual, $\MM{\Gamma}{\alpha}$, $\TT{V}{v}$ stand for respectively
$\M{\Gamma \cup \lbrace \alpha \rbrace}$, $\T{V \cup \lbrace v \rbrace}$, etc.

\begin{remark}
The notations $\M{\Gamma}$, $\T{V}$, etc. should contain the semantic structure on which
they are based. To increase readability, we will omit it. There will never be any ambiguity.
We will omit similar things with other notations in the sequel, for the same reason.
\end{remark}
A semantic structure defines a basic consequence relation:

\begin{notation}
We denote by $\cal P$ the power set operator.
\\
Let $\langle {\cal F}, {\cal V}, \models \rangle$ be a semantic structure.
\\
We denote by $\vdash$ the relation on ${\cal P}({\cal F}) \times {\cal F}$
such that $\forall \: \Gamma \subseteq {\cal F}$, $\forall \: \alpha \in {\cal F}$,
$$\Gamma \vdash \alpha\; \textrm{iff}\; \M{\Gamma} \subseteq \M{\alpha}.$$
Let $\mid\!\sim$ be a relation on ${\cal P}({\cal F}) \times {\cal F}$. Then,
\\
$\C{\mid\!\sim}{\Gamma} := \lbrace \alpha \in {\cal F} : \Gamma \mid\!\sim \alpha \rbrace$.
\\
Suppose $\cal L$ is a language, $\neg$ a unary connective of $\cal L$,
$\cal F$ the set of all wffs of $\cal L$, and $\Gamma \subseteq {\cal F}$.
\\
Then, we say that $\Gamma$ is {\it consistent} iff $\forall \: \alpha \in {\cal F}$, $\Gamma \not\vdash \alpha$ or $\Gamma \not\vdash \neg\alpha$.
\end{notation}
The following trivial facts hold, we will use them implicitly in the sequel:
\begin{remark}
Let $\langle {\cal F}, {\cal V}, \models \rangle$ be a semantic structure and $\Gamma, \Delta \subseteq {\cal F}$. Then:
\\
$\M{\Gamma, \Delta} = \M{\Gamma} \cap \M{\Delta}$;
\\
$\C{\vdash}{\Gamma} = \T{\M{\Gamma}}$;
\\
$\M{\Gamma} = \M{\C{\vdash}{\Gamma}}$;
\\
$\Gamma \subseteq \C{\vdash}{\Delta}$ iff $\C{\vdash}{\Gamma} \subseteq \C{\vdash}{\Delta}$
iff $\M{\Delta} \subseteq \M{\Gamma}$.
\end{remark}
Sometimes, we will need to make some of the following assumptions about a semantic structure:

\begin{definition}
Let $\langle {\cal F}, {\cal V}, \models \rangle$ be a semantic structure.
\\
The define the following assumptions:
\begin{description}
\item[$(A0)$] $\M{{\cal F}} = \emptyset$;
\item[$(A1)$] $\cal V$ is finite.
\end{description}
Suppose $\cal L$ is a language, $\neg$ a unary connective of $\cal L$,
and $\cal F$ the set of all wffs of $\cal L$. Then, define:
\begin{description}
\item[$(A2)$] $\forall \: \Gamma \subseteq {\cal F}$, $\forall \: \alpha \in {\cal F}$, if $\alpha \not\in \T{\M{\Gamma}}$ and $\neg\alpha \not\in \T{\M{\Gamma}}$, then $\M{\Gamma} \cap \M{\alpha} \not\subseteq \M{\neg\alpha}$.
\end{description}
Suppose $\vee$ and $\wedge$ are binary connectives of $\cal L$. Then, define:
\begin{description}
\item[$(A3)$] $\forall \: \alpha, \beta \in {\cal F}$, we have:
\hfill \\
$\M{\alpha \vee \beta} = \M{\alpha} \cup \M{\beta}$;
\hfill \\
$\M{\alpha \wedge \beta} = \M{\alpha} \cap \M{\beta}$;
\hfill \\
$\M{\neg\neg\alpha} = \M{\alpha}$;
\hfill \\
$\M{\neg(\alpha \vee \beta)} = \M{\neg\alpha \wedge \neg\beta}$;
\hfill \\
$\M{\neg(\alpha \wedge \beta)} = \M{\neg\alpha \vee \neg\beta}$.
\end{description}
\end{definition}
Clearly, those assumptions are satisfied by classical semantic structures, i.e.
structures where $\cal F$, $\cal V$, and $\models$ are classical.
In addition, we will see, in Sections~\ref{PIVfourframework} and \ref{PIVj3framework},
that they are also satisfied by certain many-valued semantic structures.

%%%%%%%%%%%%%%%%%%%%%%%
\subsubsection{The semantic structure defined by $\cal FOUR$} \label{PIVfourframework}

The logic $\cal FOUR$ was introduced by N. Belnap in \cite{Belnap2, Belnap1}.
This logic is useful to deal with inconsistent information.
Several presentations are possible, depending on the language under consideration.
For the needs of the present paper, a classical propositional language will be sufficient.
The logic has been investigated intensively in e.g. \cite{ArieliAvron2, ArieliAvron3, ArieliAvron1},
where richer languages, containing an implication connective $\supset$ (first introduced by A. Avron \cite{Avron1}),
were considered.

\begin{notation}
We denote by $\cal A$ a set of propositional symbols (or atoms).
\hfill \\
We denote by ${\cal L}_c$ the classical propositional language containing $\cal A$,
the usual constants $false$ and $true$, and the usual connectives $\neg$, $\vee$, and $\wedge$.
\hfill \\
We denote by ${\cal F}_c$ the set of all wffs of ${\cal L}_c$.
\end{notation}
We briefly recall a meaning for the logic $\cal FOUR$
(more details can be found in  \cite{CarnielliLimaMarques1, Belnap2, Belnap1}).
Consider a system in which there are, on the one hand, sources of information
and, on the other hand, a processor that listens to them.
The sources provide information about the atoms only,
not about the compound formulas.
For each atom $p$, there are exactly four possibilities:
either the processor is informed (by the sources, taken as a whole) that $p$ is true; or he is informed that $p$ is false;
or he is informed of both; or he has no information about $p$.

\begin{notation}
Denote by $0$ and $1$ the classical truth values and define:
\hfill \\
${\bf f} := \lbrace 0 \rbrace$;\qquad ${\bf t} := \lbrace 1 \rbrace$;\qquad
$\top := \lbrace 0, 1 \rbrace$;\qquad $\bot := \emptyset$.
\end{notation}
The global information given by the sources to the processor can be modelled by
a function $s$ from ${\cal A}$ to $\lbrace {\bf f}, {\bf t}, \top, \bot  \rbrace$.
Intuitively, $1 \in s(p)$ means the processor is informed that $p$ is true,
whilst $0 \in s(p)$ means he is informed that $p$ is false.

Then, the processor naturally builds information about the compound formulas from $s$.
Before he starts to do so, the situation can be
be modelled by a function $v$ from ${\cal F}_c$ to $\lbrace {\bf f}, {\bf t}, \top, \bot  \rbrace$
which agrees with $s$ about the atoms
and which assigns $\bot$ to all compound formulas.
Now, take $p$ and $q$ in $\cal A$ and suppose $1 \in v(p)$ or $1 \in v(q)$.
Then, the processor naturally adds $1$ to $v(p\vee q)$.
Similarly, if $0 \in v(p)$ and $0 \in v(q)$, then he adds $0$ in $v(p \vee q)$.
Of course, such rules hold for $\neg$ and $\wedge$ too.

Suppose all those rules are applied recursively to all compound formulas.
Then, $v$ represents the ``full'' (or developed) information given by the sources to the processor.
Now, the valuations of the logic $\cal FOUR$ can be defined as exactly those functions
that can be built
in this manner (i.e. like $v$) from some information sources. More formally,

\begin{definition}
We say that $v$ is a {\it four-valued valuation}
iff $v$ is a function from ${\cal F}_c$ to $\lbrace {\bf f}, {\bf t}, \top, \bot  \rbrace$
such that $v(true) = {\bf t}$, $v(false) = {\bf f}$ and $\forall \: \alpha, \beta \in {\cal F}_c$,
\hfill \\
$1 \in v(\neg\alpha)$ iff $0 \in v(\alpha)$;
\hfill \\
$0 \in v(\neg\alpha)$ iff $1 \in v(\alpha)$;
\hfill \\
$1 \in v(\alpha \vee \beta)$ iff $1 \in v(\alpha)$ or $1 \in v(\beta)$;
\hfill \\
$0 \in v(\alpha \vee \beta)$ iff $0 \in v(\alpha)$ and $0 \in v(\beta)$;
\hfill \\
$1 \in v(\alpha \wedge \beta)$ iff $1 \in v(\alpha)$ and $1 \in v(\beta)$;
\hfill \\
$0 \in v(\alpha \wedge \beta)$ iff $0 \in v(\alpha)$ or $0 \in v(\beta)$.
\hfill \\
We denote by ${\cal V}_4$ the set of all four-valued valuations.
\end{definition}
The definition may become more accessible if we see the four-valued valuations
as those functions that satisfy Tables 1, 2, and 3 below:
\begin{center}
\begin{tabular}{cc}
$v(\alpha)$ & $v(\neg\alpha)$\\
\hline
\multicolumn{1}{|c|}{$\bf f$} & \multicolumn{1}{|c|}{$\bf t$}\\
\multicolumn{1}{|c|}{$\bf t$} & \multicolumn{1}{|c|}{$\bf f$} \\
\multicolumn{1}{|c|}{$\top$} & \multicolumn{1}{|c|}{$\top$} \\
\multicolumn{1}{|c|}{$\bot$} & \multicolumn{1}{|c|}{$\bot$} \\
\hline
\multicolumn{2}{c}{Table 1.}
\end{tabular}
\begin{tabular}{cccccc}
 & & \multicolumn{4}{c}{$v(\beta)$}\\
 \cline{3-6}
 & & \multicolumn{1}{|c}{$\bf f$} & \multicolumn{1}{c}{$\bf t$} & \multicolumn{1}{c}{$\top$} & \multicolumn{1}{c|}{$\bot$}  \\
\cline{2-6}
 \multirow{4}{*}{$v(\alpha)$} &  \multicolumn{1}{|c|}{$\bf f$}  & \multicolumn{1}{|c}{$\bf f$}    & \multicolumn{1}{c}{$\bf t$}
 & \multicolumn{1}{c}{$\top$} & \multicolumn{1}{c|}{$\bot$}        \\
 & \multicolumn{1}{|c|}{$\bf t$}  & \multicolumn{1}{|c}{$\bf t$}    & \multicolumn{1}{c}{$\bf t$} 
 & \multicolumn{1}{c}{$\bf t$} & \multicolumn{1}{c|}{$\bf t$}        \\
& \multicolumn{1}{|c|}{$\top$} & \multicolumn{1}{|c}{$\top$} & \multicolumn{1}{c}{$\bf t$}
& \multicolumn{1}{c}{$\top$} & \multicolumn{1}{c|}{$\bf t$}      \\
& \multicolumn{1}{|c|}{$\bot$} & \multicolumn{1}{|c}{$\bot$} & \multicolumn{1}{c}{$\bf t$}
& \multicolumn{1}{c}{$\bf t$} & \multicolumn{1}{c|}{$\bot$}       \\
\cline{2-6}
&  & \multicolumn{4}{c}{$v(\alpha \vee \beta)$}\\
\multicolumn{6}{c}{Table 2.}
\end{tabular}
\begin{tabular}{cccccc}
 & & \multicolumn{4}{c}{$v(\beta)$}\\
 \cline{3-6}
 & & \multicolumn{1}{|c}{$\bf f$} & \multicolumn{1}{c}{$\bf t$} & \multicolumn{1}{c}{$\top$} & \multicolumn{1}{c|}{$\bot$}  \\
\cline{2-6}
 \multirow{4}{*}{$v(\alpha)$} &  \multicolumn{1}{|c|}{$\bf f$}  & \multicolumn{1}{|c}{$\bf f$}    & \multicolumn{1}{c}{$\bf f$} 
  & \multicolumn{1}{c}{$\bf f$} & \multicolumn{1}{c|}{$\bf f$}        \\
 & \multicolumn{1}{|c|}{$\bf t$}  & \multicolumn{1}{|c}{$\bf f$}    & \multicolumn{1}{c}{$\bf t$}    
 & \multicolumn{1}{c}{$\top$} & \multicolumn{1}{c|}{$\bot$}        \\
& \multicolumn{1}{|c|}{$\top$} & \multicolumn{1}{|c}{$\bf f$} & \multicolumn{1}{c}{$\top$}
& \multicolumn{1}{c}{$\top$} & \multicolumn{1}{c|}{$\bf f$}      \\
& \multicolumn{1}{|c|}{$\bot$} & \multicolumn{1}{|c}{$\bf f$} & \multicolumn{1}{c}{$\bot$}
 & \multicolumn{1}{c}{$\bf f$} & \multicolumn{1}{c|}{$\bot$}       \\
\cline{2-6}
& & \multicolumn{4}{c}{$v(\alpha \wedge \beta)$}\\
\multicolumn{6}{c}{Table 3.}
\end{tabular}
\end{center}
In the logic $\cal FOUR$, a formula $\alpha$ is considered to be satisfied iff the processor is informed
that it is true (it does not matter whether he is also informed that $\alpha$ is false).

\begin{notation}
We denote by $\models_4$ the relation on ${\cal V}_4 \times {\cal F}_c$ such that
$\forall \: v \in {\cal V}_4$, $\forall \: \alpha \in {\cal F}_c$, we have
\hfill \\
$v \models_4 \alpha$ iff $1 \in v(\alpha)$.
\end{notation}
Proof systems for the consequence relation $\vdash$
based on the semantic structure $\langle {\cal F}_c, {\cal V}_4, \models_4 \rangle$
(i.e. the semantic structure defined by $\cal FOUR$)
can be found in e.g. \cite{ArieliAvron2, ArieliAvron3, ArieliAvron1}.

Note that the $\cal FOUR$ semantic structure satisfies $(A0)$ and $(A3)$.
In addition, if $\cal A$ is finite, then $(A1)$ is also satisfied.
However, $(A2)$ is not satisfied by this structure.
In Section~\ref{PIVj3framework}, we turn to a many-valued semantic structure which satisfies $(A2)$.

%%%%%%%%%%%%%%%%%%%%%%%%%%%%%%%%%%%%%%%%%%%%%%%%%%
\subsubsection{The semantic structure defined by $J_3$} \label{PIVj3framework}

The logic $J_3$ was introduced in \cite{DottavianoDaCosta1} to answer a question posed in 1948 by S. Ja\'{s}kowski, who was interested in systematizing theories capable of containing contradictions, especially if they occur in dialectical reasoning. The step from informal reasoning under contradictions and formal reasoning with databases and information was done in \cite{CarnielliMarcosAmo1} (also specialized for real database models in \cite{CarnielliMarcosAmo2}), where another formulation of $J_3$ called {\bf LFI1} was introduced, and its first-order version, semantics and proof theory were studied in detail.
Investigations of $J_3$ have also been made in e.g. \cite{Avron1},
where richer languages than our ${\cal L}_c$ were considered.

The valuations of the logic $J_3$ can be given the same meaning as those of the logic $\cal FOUR$,
except that the consideration is restricted to those sources which always give some information about an atom.
More formally,

\begin{definition}
We say that $v$ is a {\it three-valued valuation} iff $v$ is a function from ${\cal F}_c$ to $\lbrace {\bf f}, {\bf t}, \top \rbrace$
such that $v(true) = {\bf t}$, $v(false) = {\bf f}$ and $\forall \: \alpha, \beta \in {\cal F}_c$,
\hfill \\
$1 \in v(\neg\alpha)$ iff $0 \in v(\alpha)$;
\hfill \\
$0 \in v(\neg\alpha)$ iff $1 \in v(\alpha)$;
\hfill \\
$1 \in v(\alpha \vee \beta)$ iff $1 \in v(\alpha)$ or $1 \in v(\beta)$;
\hfill \\
$0 \in v(\alpha \vee \beta)$ iff $0 \in v(\alpha)$ and $0 \in v(\beta)$;
\hfill \\
$1 \in v(\alpha \wedge \beta)$ iff $1 \in v(\alpha)$ and $1 \in v(\beta)$;
\hfill \\
$0 \in v(\alpha \wedge \beta)$ iff $0 \in v(\alpha)$ or $0 \in v(\beta)$.
\hfill \\
We denote by ${\cal V}_3$ the set of all three-valued valuations.
\end{definition}
As previously, the definition may become more accessible
if we see the three-valued valuations as those functions that satisfy Tables 4, 5, and 6 below:

\begin{center}
\begin{tabular}{cc}
$v(\alpha)$ & $v(\neg\alpha)$\\
\hline
\multicolumn{1}{|c|}{$\bf f$} & \multicolumn{1}{|c|}{$\bf t$}\\
\multicolumn{1}{|c|}{$\bf t$} & \multicolumn{1}{|c|}{$\bf f$} \\
\multicolumn{1}{|c|}{$\top$} & \multicolumn{1}{|c|}{$\top$} \\
\hline
\multicolumn{2}{c}{Table 4.}
\end{tabular}
\begin{tabular}{ccccc}
 & & \multicolumn{3}{c}{$v(\beta)$}\\
 \cline{3-5}
 & & \multicolumn{1}{|c}{$\bf f$} & \multicolumn{1}{c}{$\bf t$} & \multicolumn{1}{c|}{$\top$} \\
\cline{2-5}
 \multirow{3}{*}{$v(\alpha)$} &  \multicolumn{1}{|c|}{$\bf f$}  & \multicolumn{1}{|c}{$\bf f$}    & \multicolumn{1}{c}{$\bf t$}    & \multicolumn{1}{c|}{$\top$}    \\
 & \multicolumn{1}{|c|}{$\bf t$}  & \multicolumn{1}{|c}{$\bf t$}    & \multicolumn{1}{c}{$\bf t$}    & \multicolumn{1}{c|}{$\bf t$}    \\
& \multicolumn{1}{|c|}{$\top$} & \multicolumn{1}{|c}{$\top$} & \multicolumn{1}{c}{$\bf t$} & \multicolumn{1}{c|}{$\top$}  \\
\cline{2-5}
&  & \multicolumn{3}{c}{$v(\alpha \vee \beta)$}\\
\multicolumn{5}{c}{Table 5.}
\end{tabular}
\begin{tabular}{ccccc}
 & & \multicolumn{3}{c}{$v(\beta)$}\\
 \cline{3-5}
 & & \multicolumn{1}{|c}{$\bf f$} & \multicolumn{1}{c}{$\bf t$} & \multicolumn{1}{c|}{$\top$} \\
\cline{2-5}
 \multirow{3}{*}{$v(\alpha)$} &  \multicolumn{1}{|c|}{$\bf f$}  & \multicolumn{1}{|c}{$\bf f$}   & \multicolumn{1}{c}{$\bf f$}    & \multicolumn{1}{c|}{$\bf f$}    \\
 & \multicolumn{1}{|c|}{$\bf t$}  & \multicolumn{1}{|c}{$\bf f$}    & \multicolumn{1}{c}{$\bf t$}    & \multicolumn{1}{c|}{$\top$}    \\
& \multicolumn{1}{|c|}{$\top$} & \multicolumn{1}{|c}{$\bf f$} & \multicolumn{1}{c}{$\top$} & \multicolumn{1}{c|}{$\top$}  \\
\cline{2-5}
& & \multicolumn{3}{c}{$v(\alpha \wedge \beta)$}\\
\multicolumn{5}{c}{Table 6.}
\end{tabular}
\end{center}
We turn to the satisfaction relation.

\begin{notation}
We denote by $\models_3$ the relation on ${\cal V}_3 \times {\cal F}_c$ such that
$\forall \: v \in {\cal V}_3$, $\forall \: \alpha \in {\cal F}_c$, we have
\hfill \\
$v \models_3 \alpha$ iff $1 \in v(\alpha)$.
\end{notation}
Proof systems for the consequence relation $\vdash$
based on the semantic structure $\langle {\cal F}_c, {\cal V}_3, \models_3 \rangle$
(i.e. the semantic structure defined by $J_3$) have been provided in e.g.
\cite{Avron1, DottavianoDaCosta1} and chapter IX of \cite{Epstein1}.
The $J_3$ structure satisfies $(A0)$, $(A3)$ and $(A2)$.
In addition, if $\cal A$ is finite, then it satisfies $(A1)$ too.

%%%%%%%%%%%%%%%%%%
\subsection{Choice functions} \label{PIVchoicefun}

%%%%%%%%%%%%%%%%%%
\subsubsection{Definitions and properties}

In many situations, an agent has some way to choose
in any set of valuations $V$, those elements that are preferred (the bests, the more normal, etc.),
not necessarily in the absolute sense, but when the valuations in $V$ are the only ones under consideration.
In Social Choice, this is modelled by choice functions
\cite{Chernoff1, Arrow1, Sen1, AizermanMalishevski1, Lehmann1, Lehmann2}.

\begin{definition}
Let $\cal V$ be a set, ${\bf V} \subseteq {\cal P}({\cal V})$, ${\bf W} \subseteq {\cal P}({\cal V})$, and
$\mu$ a function from $\bf V$ to $\bf W$.
\hfill \\
We say that $\mu$ is a {\it choice function} iff $\forall \: V \in {\bf V}$, $\mu(V) \subseteq V$.
\end{definition}
Several properties for choice functions have been put in evidence by
researchers in Social Choice.
For the sake of completeness, we present two important ones though we will not investigate them in the present paper (a better presentation can be found in \cite{Lehmann2}).

Suppose $W$ is a set of valuations, $V$ is a subset of $W$,
and $v \in V$ is a preferred valuation of $W$.
Then, a natural requirement is that $v$ is a preferred valuation of $V$.
Indeed, in many situations, the larger a set is, the harder it is to be a preferred element of it,
and he who can do the most can do the least.
This property appears in \cite{Chernoff1} and has been given the name Coherence
in \cite{Moulin1}.

We turn to the second property.
Suppose $W$ is a set of valuations, $V$ is a subset of $W$, and
suppose all the preferred valuations of $W$ belong to $V$.
Then, they are expected to include all the preferred valuations of $V$.
The importance of this property has been put in evidence by \cite{Aizerman1, AizermanMalishevski1}
and has been given the name Local Monotonicity in e.g. \cite{Lehmann2}.

In~\cite{Schlechta1}, Schlechta showed that Coherence and Local Monotonicity
characterize those choice functions that can be defined by a binary preference relation
on states labelled by valuations (in the style of e.g.
\cite{KrausLehmannMagidor1}).

Now, we turn to properties relevant for the paper, i.e. properties which characterize those choice functions
that can be defined by a pivot
(in the style of e.g. D. Makinson \cite{Makinson1, Makinson2}). A pivot is a fixed subset of valuations
which are considered to be the important ones in the absolute sense.
Details will be given in Section~\ref{PIVPivots}.

\begin{definition}
Let $\cal V$ be a set,
${\bf V} \subseteq {\cal P}({\cal V})$, ${\bf W} \subseteq {\cal P}({\cal V})$, and
$\mu$ a choice function from $\bf V$ to $\bf W$.
\\
We say that $\mu$ is {\it strongly coherent} (SC) iff $\forall \: V, W \in {\bf V}$,
$$\mu(W) \cap V \subseteq \mu(V).$$
Suppose $\langle {\cal F}, {\cal V}, \models \rangle$ is a semantic structure.
\\
We say that $\mu$ is {\it definability preserving} (DP) iff
$$\forall \: V \in {\bf V} \cap {\bf D},\; \mu(V) \in {\bf D}.$$
In addition, suppose ${\cal V} \in {\bf V}$.
\\
We say that $\mu$ is {\it universe-codefinable} (UC) iff
$${\cal V} \setminus \mu({\cal V}) \in {\bf D}.$$
\end{definition}
Definability Preservation has been put in evidence first in \cite{Schlechta2}.
One of its advantages
is that when the choice functions under consideration
satisfy it, we will provide characterizations with purely syntactic conditions.
To the author knowledge, Strong Coherence and Universe-codefinability are
first introduced in the present paper.
An advantage of Universe-codefinability is that it provides
a link with $X$-logics \cite{ForgetRischSiegel1}. We will see it in Section~\ref{PIVsectioncodefisclosed}.

Now, we turn to a last property:

\begin{definition}
Suppose $\cal L$ is a language, $\neg$ a unary connective of $\cal L$,
$\cal F$ the set of all wffs of $\cal L$, $\langle {\cal F}, {\cal V}, \models \rangle$
a semantic structure, ${\bf V} \subseteq {\cal P}({\cal V})$, ${\bf W} \subseteq {\cal P}({\cal V})$, and
$\mu$ be a choice function from $\bf V$ to $\bf W$.
\\
We say that $\mu$ is {\it coherency preserving} (CP) iff
$$\forall \: V \in {\bf V} \cap {\bf C},\; \mu(V) \in {\bf C}.$$
\end{definition}
To the author knowledge, Coherency Preservation has been first introduced in \cite{Bennaim1}.
An advantage of it is that when the choice functions under consideration satisfy it, we will not need
to assume $(A2)$ to show our characterizations (in the discriminative case).

%%%%%%%%%%%%
\subsubsection{Pivots} \label{PIVPivots}

Suppose some valuations are considered to be the important ones in the absolute sense and
collect them in a set $\cal I$, called a pivot. Then, $\cal I$ defines naturally a choice
function $\mu_{\cal I}$ which chooses in any set of valuations, simply those elements which belong to $\cal I$.
More formally,

\begin{definition}
Let $\cal V$ be a set.
\\
We say that $\cal I$ is a pivot on $\cal V$ iff ${\cal I} \subseteq {\cal V}$.
\\
Let $\cal I$ be a pivot on $\cal V$.
\\
We denote by $\mu_{\cal I}$ the function from ${\cal P}({\cal V})$ to ${\cal P}({\cal V})$ such that
$\forall \: V \subseteq {\cal V}$,
$$\mu_{\cal I}(V) = V \cap {\cal I}.$$
\end{definition}
Pivots have been investigated extensively by D.~Makinson in \cite{Makinson1, Makinson2}.
In the present section, we show that the properties of Strong Coherence, Definability Preservation, and Universe-codefinability characterize those choice functions that can be defined by a pivot. More precisely:

\begin{proposition} \label{PIVmupp}
Let $\cal V$ be a set, ${\bf V}, {\bf W} \subseteq {\cal P}({\cal V})$,
and $\mu$ a choice function from $\bf V$ to $\bf W$. Then:
\begin{description}
\item[$(0)$] $\mu$ is SC iff there exists a pivot $\cal I$ on ${\cal V}$ such that
$\forall \: V \in {\bf V}$, $\mu(V) = \mu_{\cal I}(V)$.
\end{description}
Suppose $\langle {\cal F}, {\cal V}, \models \rangle$ is a semantic structure and ${\cal V} \in {\bf V}$. Then:
\begin{description}
\item[$(1)$] $\mu$ is SC and DP iff there exists a pivot $\cal I$ on ${\cal V}$ such that ${\cal I} \in {\bf D}$
and $\forall \: V \in {\bf V}$, $\mu(V) = \mu_{\cal I}(V)$;
\item[$(2)$] $\mu$ is SC and UC iff there exists a pivot $\cal I$ on ${\cal V}$ such that
${\cal V} \setminus {\cal I} \in {\bf D}$ and $\forall \: V \in {\bf V}$, $\mu(V) = \mu_{\cal I}(V)$.
\end{description}
\end{proposition}

\begin{proof} {\it Proof of $(0)$}. Direction: ``$\rightarrow$''.
\\
Let ${\cal I} = \lbrace v \in {\cal V} : \exists \: V \in {\bf V}$, $v \in \mu(V) \rbrace$ and suppose $V \in {\bf V}$.
\\
If $v \in \mu(V)$, then $v \in V$ and, by definition of $\cal I$, $v \in {\cal I}$.
Consequently, $\mu(V) \subseteq V \cap {\cal I}$.
\\
If $v \in V \cap {\cal I}$, then $\exists \: W \in {\bf V}$, $v \in \mu(W)$, thus, by SC, $v \in \mu(W) \cap V \subseteq \mu(V)$.
\\
Consequently, $V \cap {\cal I} \subseteq \mu(V)$.

Direction: ``$\leftarrow$''.
\\
There exists ${\cal I} \subseteq {\cal V}$ such that $\forall \: V \in {\bf V}$, $\mu(V) = V \cap {\cal I}$.
\\
We show that $\mu$ satisfies SC.
\\
Let $V, W \in {\bf V}$.
Then, $\mu(W) \cap V = W \cap {\cal I} \cap V \subseteq {\cal I} \cap V = \mu(V)$.
\\ \\
{\it Proof of $(1)$}. Direction: ``$\rightarrow$''.
\\
Take the same $\cal I$ as for $(0)$. Then, by verbatim the same proof,
$\forall \: V \in {\bf V}$, $\mu(V) = V \cap {\cal I}$.
\\
It remains to show that ${\cal I} \in {\bf D}$.
\\
As $\M{\emptyset} = {\cal V}$, ${\cal V} \in {\bf D}$.
Thus, as $\mu$ is DP, $\mu({\cal V}) \in {\bf D}$.
But, $\mu({\cal V}) = {\cal V} \cap {\cal I} = {\cal I}$.

Direction: ``$\leftarrow$''.
\\
Verbatim the proof of $(0)$, except that in addition ${\cal I} \in {\bf D}$.
\\
We show that $\mu$ is DP.
Let $V \in {\bf V} \cap {\bf D}$.
\\Then, $\exists \: \Gamma \subseteq {\cal F}$, $\M{\Gamma} = V$.
Similarly, as ${\cal I} \in {\bf D}$, $\exists \: \Delta \subseteq {\cal F}$, $\M{\Delta} = {\cal I}$.
\\
Therefore, $\mu(V) = V \cap {\cal I} = \M{\Gamma} \cap \M{\Delta} = \M{\Gamma \cup \Delta} \in {\bf D}$.
\\ \\
{\it Proof of $(2)$}. Direction: ``$\rightarrow$''.
\\
Take the same $\cal I$ as for $(0)$. Then, by verbatim the same proof,
$\forall \: V \in {\bf V}$, $\mu(V) = V \cap {\cal I}$.
\\
It remains to show ${\cal V} \setminus {\cal I} \in {\bf D}$.
As $\mu$ is UC, ${\cal V} \setminus \mu({\cal V}) \in {\bf D}$.
But, ${\cal V} \setminus \mu({\cal V}) = {\cal V} \setminus ({\cal V} \cap {\cal I}) =
{\cal V} \setminus {\cal I}$.

Direction: ``$\leftarrow$''.
\\
Verbatim the proof of $(0)$, except that in addition ${\cal V} \setminus {\cal I} \in {\bf D}$.
\\
We show that $\mu$ is UC: ${\cal V} \setminus \mu({\cal V}) = {\cal V} \setminus ({\cal V} \cap {\cal I}) =
{\cal V} \setminus {\cal I} \in {\bf D}$.\qed
\end{proof}

%%%%%%%%%%%%%%%%%%%%%%%
\subsection{Pivotal(-discriminative) consequence relations} \label{PIVintroprefCR}

%%%%%%%%%%%%%%%%%%%%%%%
\subsubsection{Definitions}

Suppose we are given a semantic structure and a choice function $\mu$ on the valuations.
Then, it is natural to conclude a formula $\alpha$ from a set of formulas $\Gamma$ iff
every model for $\Gamma$ chosen by $\mu$ is a model for $\alpha$.
More formally:

\begin{definition}
Let $\langle {\cal F}, {\cal V}, \models \rangle$ be a semantic structure and $\mid\!\sim$ a relation
on ${\cal P}({\cal F}) \times {\cal F}$.
\\
We say that $\mid\!\sim$ is a {\it pivotal consequence relation} iff there exists
a SC choice function $\mu$ from $\bf D$ to ${\cal P}({\cal V})$ such that
$\forall \: \Gamma \subseteq {\cal F}$, $\forall \: \alpha \in {\cal F}$,
$$
\Gamma \mid\!\sim \alpha\; \textrm{iff}\; \mu(\M{\Gamma}) \subseteq \M{\alpha}
$$
In addition, if $\mu$ is DP, CP, etc., then so is $\mid\!\sim$.
\end{definition}
We called these relations ``pivotal'' because, in the light of Proposition~\ref{PIVmupp}, they
can be defined equivalently with pivots, instead of SC choice functions.
Their importance has been put in evidence by D.~Makinson in e.g. \cite{Makinson1, Makinson2},
where he showed that they constitute easy conceptual passage
from basic to plausible non-monotonic consequence relations.
Indeed, they are perfectly monotonic but already display some of the distinctive
features (i.e. the choice functions) of plausible non-monotonic relations.
Note that pivotal (resp. DP pivotal) consequence relations
correspond to Makinson's pivotal-valuation (resp. pivotal-assumption) relations.
We will give an example of how they can be used to draw plausible conclusions from
incomplete information
in Section~\ref{PIVclassicalexample}.

Moreover, if a many-valued semantic structure is considered,
they lead to rational and non-trivial
conclusions is spite of the presence of contradictions and are thus useful to treat
both incomplete and inconsistent information. However, they will not satisfy
the Disjunctive Syllogism.
We will give an example with the $\cal FOUR$ semantic structure in Section~\ref{PIVfourexample}.

Characterizations of pivotal consequence relations, valid in classical frameworks, can be found in the literature.
For instance, the following result appears to be part of folklore for decades:
DP pivotal consequence relations correspond precisely to those supraclassical closure operations that are
compact and satisfy Disjunction in the premisses.
For more details see e.g. \cite{Rott1, Makinson1, Makinson2}.

Now, we turn to a qualified version of pivotal consequence.
It captures the idea that the contradictions in the conclusions should be rejected.

\begin{definition}
Let $\cal L$ be a language, $\neg$ a unary connective of $\cal L$, $\cal F$ the set of all wffs of $\cal L$,
$\langle {\cal F}, {\cal V}, \models \rangle$ a semantic structure, and $\mid\!\sim$ a relation
on ${\cal P}({\cal F}) \times {\cal F}$.
\\
We say that $\mid\!\sim$ is a {\it pivotal-discriminative consequence relation} iff there exists
a SC choice function $\mu$ from $\bf D$ to ${\cal P}({\cal V})$ such that
$\forall \: \Gamma \subseteq {\cal F}$, $\forall \: \alpha \in {\cal F}$,
$$
\Gamma \mid\!\sim \alpha\; \textrm{iff}\; \mu(\M{\Gamma}) \subseteq \M{\alpha}
\;\textrm{and}\; \mu(\M{\Gamma}) \not\subseteq \M{\neg\alpha}
$$
In addition, if $\mu$ is DP, CP, etc., then so is $\mid\!\sim$.
\end{definition}
If a classical semantic structure is considered, the discriminative version does not bring something really new.
Indeed, the only difference will be to conclude nothing instead of everything in the face of inconsistent information.
On the other hand, with a many-valued structure,
the conclusions are rational even from inconsistent information.
The discriminative version will then reject the contradictions in the conclusions,
rendering the latter all the more rational.

%%%%%%%%%%%%%%%%%%%%%%%%%%%%%%%%%%%%
\subsubsection{Example in the classical framework} \label{PIVclassicalexample}

Let $\cal L$ be a classical propositional language of which the atoms are
$r$, $q$, and $p$.
Intuitively, $r$ means Nixon is a republican, $q$ means Nixon is a quaker, and $p$ means Nixon is a pacifist.
Let $\cal F$ be the set of all wffs of $\cal L$,
$\cal V$ the set of all classical two-valued valuations of $\cal L$, and $\models$
the classical satisfaction relation for these objects.
Then, ${\cal V}$ is the set of the 8 following valuations: $v_0$, $v_1$, $v_2$, $v_3$, $v_4$, $v_5$, $v_6$, and $v_7$,
which are defined in the obvious way by the following table:
$$
\begin{array}{|c||c|c|c|}
\hline
               & r & q & p \\
\hline v_0 & 0 & 0 & 0 \\
\hline v_1 & 0 & 0 & 1 \\
\hline v_2 & 0 & 1 & 0 \\
\hline v_3 & 0 & 1 & 1 \\
\hline v_4 & 1 & 0 & 0 \\
\hline v_5 & 1 & 0 & 1 \\
\hline v_6 & 1 & 1 & 0 \\
\hline v_7 & 1 & 1 & 1 \\
\hline
\end{array}
$$
Now, consider the class of all republicans and the class of all quakers.
Consider that a republican is normal iff he is not a pacifist and that a quaker is normal iff he is a pacifist.
And, consider that a valuation is negligible iff (in it)
Nixon is a non-normal individual of some class.
Then, collect the non-negligible valuations in a pivot $\cal I$.
More formally:
$${\cal I} = \lbrace v \in {\cal V} : \textrm{if}\; v \models r, \;\textrm{then}\; v \models \neg p;\;
\textrm{and if}\; v \models q, \;\textrm{then}\; v \models p \rbrace.$$
Finally, let $\mid\!\sim$ be the pivotal consequence relation defined by the SC choice function $\mu_{\cal I}$.

Then, $\mid\!\sim$ leads to ``jump'' to plausible conclusions from incomplete information.
For instance, $r \mid\!\sim \neg p$ and $q \mid\!\sim p$.
But, we fall into triviality if we face new information that contradict previous ``hasty'' conclusions.
For instance, $\lbrace r, p \rbrace \mid\!\sim \alpha$, $\forall \: \alpha \in {\cal L}$,
and $\lbrace q, \neg p \rbrace \mid\!\sim \alpha$, $\forall \: \alpha \in {\cal L}$.
This is the price to pay for being monotonic, whereas conclusions that are only plausible are accepted.

In addition, $\mid\!\sim$ is not paraconsistent and some sets of formulas are rendered useless because
there is no model in the pivot for them, though there are models for them.
For instance, $\lbrace q, r \rbrace \mid\!\sim \alpha$, $\forall \: \alpha \in {\cal L}$.

%%%%%%%%%%%%%%%%%%%%%%
\subsubsection{Example in the $\cal FOUR$ framework} \label{PIVfourexample}

Consider the $\cal FOUR$ semantic structure $\langle {\cal F}_c, {\cal V}_4, \models_4 \rangle$
and suppose ${\cal A} = \lbrace r, q, p \rbrace$ (these objects have been defined
in Section~\ref{PIVfourframework}). In addition, make the same considerations
about Nixon, the classes, normality, etc., as in Section~\ref{PIVclassicalexample}, except that this time
a valuation is considered to be negligible iff (in it)
the processor is informed that Nixon is an individual of some class,
but he is not informed that Nixon is a normal individual of that class.
See Section~\ref{PIVfourframework} for recalls about the sources-processor systems.
Again, collect the non-negligible valuations in a pivot $\cal I$. More formally:
$${\cal I} = \lbrace v \in {\cal V}_4 : \textrm{if}\; v \models r, \;\textrm{then}\; v \models \neg p;
\;\textrm{and if}\; v \models q, \;\textrm{then}\; v \models p \rbrace.$$
Let $\mid\!\sim$ be the pivotal consequence relation defined by the SC choice function $\mu_{\cal I}$.

Then, again $\mid\!\sim$ leads to ``jump'' to plausible conclusions from incomplete information.
For instance, $r \mid\!\sim \neg p$ and $q \mid\!\sim p$.
Moreover, though ``hasty'' conclusions are never withdrawn,
we do not fall into triviality when we face new information that contradict them.
For instance, $\lbrace r, p \rbrace \mid\!\sim p$ and $\lbrace r, p \rbrace \mid\!\sim \neg p$
and $\lbrace r, p \rbrace \mid\!\sim r$ and $\lbrace r, p \rbrace \not\mid\!\sim \neg r$.
%Here is another example: $\lbrace q, \neg p \rbrace \mid\!\sim \neg p$ and $\lbrace q, \neg p \rbrace \mid\!\sim p$
%and $\lbrace q, \neg p \rbrace \mid\!\sim q$ and $\lbrace q, \neg p \rbrace \not\mid\!\sim \neg q$.

In addition, $\mid\!\sim$ is paraconsistent.
For instance, $\lbrace p, \neg p, q \rbrace \mid\!\sim p$ and $\lbrace p, \neg p, q \rbrace \mid\!\sim \neg p$ and
$\lbrace p, \neg p, q \rbrace \mid\!\sim q$ and $\lbrace p, \neg p, q \rbrace \not\mid\!\sim \neg q$.
And, less sets of formulas are rendered useless because there is no model in the pivot for them,
though there are models for them.
For instance, this time, $\lbrace q, r \rbrace \mid\!\sim p$ and $\lbrace q, r \rbrace \mid\!\sim \neg p$ and
$\lbrace q, r \rbrace \mid\!\sim q$ and $\lbrace q, r \rbrace \not\mid\!\sim \neg q$ and $\lbrace q, r \rbrace \mid\!\sim r$ and $\lbrace q, r \rbrace \not\mid\!\sim \neg r$.

However, $\mid\!\sim$ does not satisfy the Disjunctive Syllogism.
Indeed, for instance, $\lbrace \neg r, r \vee q \rbrace \not\mid\!\sim q$.

%%%%%%%%%%%%%%%%%%%%%%%%%%%%%%%%%%%%%%%%%%%%%%%%%%%%%%
\section{Characterizations} \label{PIVcacarac}

The first contributions of the paper are characterizations (in many cases, with purely syntactic conditions)
of several families of pivotal and pivotal-discriminative consequence relations.
Sometimes, we will need to make some assumptions about the semantic structure under consideration.
However, no assumption will be needed for the two following families:
\begin{itemize}
\item the pivotal consequence relations (Section~\ref{PIVpivotal});
\item the DP pivotal consequence relations (Section~\ref{PIVdefinablepivotal}).
\end{itemize}
We will assume $(A0)$ for:
\begin{itemize}
\item the UC pivotal consequence relations (Section~\ref{PIVpivotal}).
\end{itemize}
We will need $(A3)$ and $(A1)$ for:
\begin{itemize}
\item the CP pivotal-discriminative consequence relations (Section~\ref{PIVpivotaldiscri});
\item the CP DP pivotal-discriminative consequence relations (Section~\ref{PIVdefinablepivotaldiscri}).
\end{itemize}
We will assume $(A3)$, $(A1)$, and $(A2)$ for:
\begin{itemize}
\item the pivotal-discriminative consequence relations (Section~\ref{PIVpivotaldiscri});
\item the DP pivotal-discriminative consequence relations (Section~\ref{PIVdefinablepivotaldiscri}).
\end{itemize}
We will assume $(A0)$, $(A3)$, and $(A1)$ for:
\begin{itemize}
\item the CP UC pivotal-discriminative consequence relations (Section~\ref{PIVpivotaldiscri}).
\end{itemize}
We will need $(A0)$, $(A3)$, $(A1)$, and $(A2)$ for:
\begin{itemize}
\item the UC pivotal-discriminative consequence relations (Section~\ref{PIVpivotaldiscri}).
\end{itemize}

%%%%%%%%%%%%%%%%%%%%%%%%%%
\subsection{The non-discriminative and definability preserving case} \label{PIVdefinablepivotal}

In the present section, we provide, in a general framework,
a characterization for the family of all DP pivotal consequence relations.
We will use techniques very similar to those of \cite{Bennaim1}
(see the DP and non-discriminative case).
The latter are themselves strongly inspired by the work, in a classical propositional framework, of K. Schlechta
(see Proposition~3.1 of \cite{Schlechta1}).
The idea is to get to the remarkable equality: $\mu(\M{\Gamma}) = \M{\C{\mid\!\sim}{\Gamma}}$.
Thanks to it, properties like Strong Coherence
can be easily translated in syntactic terms
(i.e. using only the language, $\vdash$, $\mid\!\sim$, etc.).

\begin{definition}
Let $\langle {\cal F}, {\cal V}, \models \rangle$ be a semantic structure and $\mid\!\sim$ be a
relation on ${\cal P}({\cal F}) \times {\cal F}$.
\\
Then, consider the following conditions: $\forall \: \Gamma, \Delta \subseteq {\cal F}$,
\begin{description}
\item[$(\mid\!\sim$$0)$] if $\C{\vdash}{\Gamma} = \C{\vdash}{\Delta}$, then
$\C{\mid\!\sim}{\Gamma} = \C{\mid\!\sim}{\Delta}$;
\item[$(\mid\!\sim$$1)$] $\C{\mid\!\sim}{\Gamma} = \C{\vdash}{\C{\mid\!\sim}{\Gamma}}$;
\item[$(\mid\!\sim$$2)$] $\Gamma \subseteq \C{\mid\!\sim}{\Gamma}$;
\item[$(\mid\!\sim$$3)$] $\C{\mid\!\sim}{\Gamma} \subseteq
\CC{\vdash}{\C{\mid\!\sim}{\Delta}}{\Gamma}$.
\end{description}
\end{definition}
Note that those conditions are purely syntactic
when there is a proof system available for $\vdash$
(which is the case with e.g. the classical, $\cal FOUR$, and $J_3$ semantic structures).

\begin{proposition} \label{PIVrepDef}
Let $\langle {\cal F}, {\cal V}, \models \rangle$ be a semantic structure and $\mid\!\sim$ be a
relation on ${\cal P}({\cal F}) \times {\cal F}$.
\\
Then, $\mid\!\sim$ is an DP pivotal consequence relation iff $\mid\!\sim$ satisfies
$(\mid\!\sim$$0)$, $(\mid\!\sim$$1)$, $(\mid\!\sim$$2)$, and $(\mid\!\sim$$3)$.
\end{proposition}

\begin{proof}
Direction: ``$\rightarrow$''.
\\
There exists an DP SC choice function $\mu$ from $\bf D$ to ${\cal P}({\cal V})$
such that $\forall \: \Gamma \subseteq {\cal F}$, $\C{\mid\!\sim}{\Gamma} = \T{\mu(\M{\Gamma})}$.
\\
We will show:
\\
$(0)$\quad $\forall \: \Gamma \subseteq {\cal F}$, $\mu(\M{\Gamma}) = \M{\C{\mid\!\sim}{\Gamma}}$;
\\
$(1)$\quad $\mid\!\sim$ satisfies $(\mid\!\sim$$0)$;
\\
$(2)$\quad $\mid\!\sim$ satisfies $(\mid\!\sim$$1)$;
\\
$(3)$\quad $\mid\!\sim$ satisfies $(\mid\!\sim$$2)$;
\\
$(4)$\quad $\mid\!\sim$ satisfies $(\mid\!\sim$$3)$.

Direction: ``$\leftarrow$''.
\\
Suppose $\mid\!\sim$ satisfies $(\mid\!\sim$$0)$, $(\mid\!\sim$$1)$, $(\mid\!\sim$$2)$, and $(\mid\!\sim$$3)$.
\\
Let $\mu$ be the function from $\bf D$ to ${\cal P}({\cal V})$ such that
$\forall \: \Gamma \subseteq {\cal F}$, $\mu(\M{\Gamma}) = \M{\C{\mid\!\sim}{\Gamma}}$.
\\
We will show:
\\
$(5)$\quad $\mu$ is well-defined;
\\
$(6)$\quad $\mu$ is a DP choice function;
\\
$(7)$\quad $\mu$ is SC;
\\
$(8)$\quad $\forall \: \Gamma \subseteq {\cal F}$, $\C{\mid\!\sim}{\Gamma} = \T{\mu(\M{\Gamma})}$.
\\ \\
{\it Proof of $(0)$.} Let $\Gamma \subseteq {\cal F}$.
As $\mu$ is DP, $\mu(\M{\Gamma}) \in {\bf D}$. Thus,
$\exists \: \Delta \subseteq {\cal F}$, $\mu(\M{\Gamma}) = \M{\Delta}$.
\\
Therefore,
$\mu(\M{\Gamma}) = \M{\Delta} = \M{\T{\M{\Delta}}} =
\M{\T{\mu(\M{\Gamma})}} = \M{\C{\mid\!\sim}{\Gamma}}$.
\\ \\
{\it Proof of $(1)$.} Let $\Gamma, \Delta \subseteq {\cal F}$ and suppose
$\C{\vdash}{\Gamma} = \C{\vdash}{\Delta}$.
\\
Then, $\M{\Gamma} = \M{\Delta}$. Thus,
$\C{\mid\!\sim}{\Gamma} = \T{\mu(\M{\Gamma})} = \T{\mu(\M{\Delta})} = \C{\mid\!\sim}{\Delta}$.
\\ \\
{\it Proof of $(2)$.} Let $\Gamma \subseteq {\cal F}$.
Then, $\C{\mid\!\sim}{\Gamma} = \T{\mu(\M{\Gamma})} =
\T{\M{\T{\mu(\M{\Gamma})}}} =
\T{\M{\C{\mid\!\sim}{\Gamma}}} =
\C{\vdash}{\C{\mid\!\sim}{\Gamma}}$.
\\ \\
{\it Proof of $(3)$.} Let $\Gamma \subseteq {\cal F}$.
Then, $\Gamma \subseteq \T{\M{\Gamma}} \subseteq \T{\mu(\M{\Gamma})} = \C{\mid\!\sim}{\Gamma}$.
\\ \\
{\it Proof of $(4)$.} Let $\Gamma, \Delta \subseteq {\cal F}$.
Then, by $(0)$ and SC,
\\
$\MM{\C{\mid\!\sim}{\Delta}}{\Gamma} = \M{\C{\mid\!\sim}{\Delta}} \cap \M{\Gamma} =
\mu(\M{\Delta}) \cap \M{\Gamma} \subseteq \mu(\M{\Gamma}) = \M{\C{\mid\!\sim}{\Gamma}}$.
\\
Therefore, by $(\mid\!\sim$$1)$, we get
$\C{\mid\!\sim}{\Gamma} = \C{\vdash}{\C{\mid\!\sim}{\Gamma}} = \T{\M{\C{\mid\!\sim}{\Gamma}}} \subseteq \T{\MM{\C{\mid\!\sim}{\Delta}}{\Gamma}}
= \CC{\vdash}{\C{\mid\!\sim}{\Delta}}{\Gamma}$.
\\ \\
{\it Proof of $(5)$.} Let $\Gamma, \Delta \subseteq {\cal F}$ and suppose
$\M{\Gamma} = \M{\Delta}$.
\\
Then, $\C{\vdash}{\Gamma} = \C{\vdash}{\Delta}$.
Thus, by $(\mid\!\sim$$0)$, $\M{\C{\mid\!\sim}{\Gamma}} =_{} \M{\C{\mid\!\sim}{\Delta}}$.
\\ \\
{\it Proof of $(6)$.} Let $\Gamma \subseteq {\cal F}$.
Then, by $(\mid\!\sim$$2)$, $\mu(\M{\Gamma}) = \M{\C{\mid\!\sim}{\Gamma}}
\subseteq \M{\Gamma}$.
\\
Consequently, $\mu$ is a choice function.
In addition, $\mu$ is obviously DP.
\\ \\
{\it Proof of $(7)$.} Let $\Gamma, \Delta \subseteq {\cal F}$.
\\
Then, by $(\mid\!\sim$$3)$, we get $\mu(\M{\Delta}) \cap \M{\Gamma} = \M{\C{\mid\!\sim}{\Delta}} \cap \M{\Gamma} = \MM{\C{\mid\!\sim}{\Delta}}{\Gamma} \subseteq
\M{\C{\mid\!\sim}{\Gamma}} = \mu(\M{\Gamma})$.
\\ \\
{\it Proof of $(8)$.} Let $\Gamma \subseteq {\cal F}$.
Then, by $(\mid\!\sim$$1)$, $\C{\mid\!\sim}{\Gamma} =
\C{\vdash}{\C{\mid\!\sim}{\Gamma}} =
\T{\M{\C{\mid\!\sim}{\Gamma}}} =
\T{\mu(\M{\Gamma})}$.\qed
\end{proof}

%%%%%%%%%%%%%%%%%%%%%%%
\subsection{The non-discriminative and not necessarily definability preserving case} \label{PIVpivotal}

In the present section, we will investigate in particular the family of all pivotal consequence relations.
Unlike in Section~\ref{PIVdefinablepivotal}, the choice functions considered here are not necessarily definability preserving.
As a consequence, we will no longer have at our disposal the remarkable equality:
$\mu(\M{\Gamma}) = \M{\C{\mid\!\sim}{\Gamma}}$.
Therefore, we cannot translate properties like Strong Coherence in syntactic terms.
Moreover, we will put in evidence, in Section~\ref{PIVsectionNoCharac},
some limits of what can be done in this area.
Approximatively, we will show, in an infinite classical framework, that
there does not exist a characterization (of the aforementioned family) made
of conditions which are universally quantified and of limited size.

We provide a solution with semi-syntactic conditions.
To do so, we will use techniques very similar to those of \cite{Bennaim1}
(see the non-DP and non-discriminative case). The latter are themselves
strongly inspired by the work of K. Schlechta (see Proposition~5.2.5 of \cite{Schlechta5}).
Technically, the idea begins by building from any function $f$, a SC choice function $\nup{f}$ such that
whenever $f$ ``covers'' some SC choice function, it necessarily covers $\nup{f}$.

\begin{definition}
Let $\cal V$ be a set, ${\bf V} \subseteq {\cal P}({\cal V})$, ${\bf W} \subseteq {\cal P}({\cal V})$ and
$f$ a function from $\bf V$ to $\bf W$.
\\
We denote by $\nup{f}$ the function from $\bf V$ to ${\cal P}({\cal V})$ such that
$\forall \: V \in {\bf V}$,
$$
\nup{f}(V) = \lbrace v \in V : \forall \: W \in {\bf V},\; \textrm{if}\; v \in W,\; \textrm{then}\; v \in f(W) \rbrace.
$$
\end{definition}

\begin{lemma} \label{PIVmufestSCf}
Let $\cal V$ be a set, ${\bf V} \subseteq {\cal P}({\cal V})$, ${\bf W} \subseteq {\cal P}({\cal V})$ and
$f$ a function from $\bf V$ to $\bf W$.
\\
Then, $\nup{f}$ is a SC choice function.
\end{lemma}

\begin{proof}
$\nup{f}$ is obviously a choice function.
We show that it satisfies Strong Coherence.
\\
Suppose the contrary, i.e. suppose $\exists \: V, W \in {\bf V}$ and
$\exists \: v \in \nupp{f}{W} \cap V$ such that $v \not\in \nupp{f}{V}$.
\\
Then, as $v \in V$ and $v \not\in \nupp{f}{V}$, we have
$\exists \: Z \in {\bf V}$, $v \in Z$, and $v \not\in f(Z)$.
\\
Therefore, simply by definition of $\nup{f}$,
$v \not\in \nupp{f}{W}$, which is impossible.\qed
\end{proof}

\begin{lemma} \label{PIVpivgen}
Let $\cal V$ be a set, ${\bf V}$, ${\bf W}$, and $\bf X$ subsets of ${\cal P}({\cal V})$,
$f$ a function from ${\bf V}$ to ${\bf W}$, and $\mu$ a SC choice function from $\bf V$ to $\bf X$
such that $\forall \: V \in {\bf V}$, $f(V) = \M{\T{\mu(V)}}$.
Then:
\begin{description}
\item[$(0)$] $\forall \: V \in {\bf V}$, $f(V) = \M{\T{\nupp{f}{V}}}$.
\end{description}
Suppose $\langle {\cal F}, {\cal V}, \models \rangle$ is a semantic structure satisfying $(A0)$,
${\bf D} \subseteq {\bf V}$, and $\mu$ is UC.
Then:
\begin{description}
\item[$(1)$] $\nupp{f}{{\cal V}} = \mu({\cal V})$.
\end{description}
\end{lemma}
\begin{proof}
{\it Proof of $(0)$.}
Suppose $V \in {\bf V}$. We show $f(V) = \M{\T{\nupp{f}{V}}}$.
\\
Case~1: $\exists \: v \in \mu(V)$, $v \not\in \nupp{f}{V}$.
\\
As $\mu(V) \subseteq V$, we have $v \in V$.
\\
Thus, by definition of $\nup{f}$,
$\exists \: W \in {\bf V}$, $v \in W$, and $v \not\in f(W) = \M{\T{\mu(W)}}
\supseteq \mu(W)$.
\\
On the other hand, as $\mu$ is SC, 
$\mu(V) \cap W \subseteq \mu(W)$. Thus, $v \in \mu(W)$, which is impossible.
\\
Case~2: $\mu(V) \subseteq \nupp{f}{V}$.
\\
Case~2.1: $\exists \: v \in \nupp{f}{V}$, $v \not\in f(V)$.
\\
Then, $\exists \: W \in {\bf V}$, $v \in W$, and $v \not\in f(W)$.
Indeed, just take $V$ itself for the choice of $W$.
\\
Therefore, by definition of $\nup{f}$,
$v \not \in \nupp{f}{V}$, which is impossible.
\\
Case~2.2: $\nupp{f}{V} \subseteq f(V)$.
\\
Then, $f(V) = \M{\T{\mu(V)}} \subseteq \M{\T{\nupp{f}{V}}} \subseteq \M{\T{f(V)}} =
\M{\T{\M{\T{\mu(V)}}}} = \M{\T{\mu(V)}} = f(V)$.
\\ \\
{\it Proof of $(1)$.} 
Direction: ``$\subseteq$''.
\\
Suppose the contrary, i.e. suppose $\exists \: v \in \nupp{f}{{\cal V}}$, $v \not\in \mu({\cal V})$.
\\
Then, $v \in {\cal V} \setminus \mu({\cal V})$.
But, as $\mu$ is UC, ${\cal V} \setminus \mu({\cal V}) \in {\bf D} \subseteq {\bf V}$.
\\
On the other hand, as $v \in \nupp{f}{{\cal V}}$, we get
$\forall \: W \in {\bf V},\; \textrm{if}\; v \in W,\; \textrm{then}\; v \in f(W)$.
\\
Therefore, $v \in f({\cal V} \setminus \mu({\cal V}))
= \M{\T{\mu({\cal V} \setminus \mu({\cal V}))}}$.
\\
But, we will show:
\\
$(1.0)$\quad $\mu({\cal V} \setminus \mu({\cal V})) = \emptyset$.
\\
Therefore, $\M{\T{\mu({\cal V} \setminus \mu({\cal V}))}} = \M{\T{\emptyset}} = \M{{\cal F}}$.
\\
But, by $(A0)$, $\M{{\cal F}} = \emptyset$.
Therefore, $v \in \emptyset$, which is impossible.

Direction: ``$\supseteq$''.
\\
Suppose the contrary, i.e. suppose $\exists \: v \in \mu({\cal V})$, $v \not\in \nupp{f}{{\cal V}}$.
\\
As $v \in {\cal V}$ and $v \not\in \nupp{f}{{\cal V}}$, we get
$\exists \: W \in {\bf V}$, $v \in W$ and $v \not\in f(W) = \M{\T{\mu(W)}} \supseteq \mu(W)$.
\\
But, as $\mu$ is SC, $\mu({\cal V}) \cap W \subseteq \mu(W)$.
Therefore, $v \in \mu(W)$, which is impossible.
\\ \\
{\it Proof of $(1.0)$.}
Suppose the contrary, i.e. suppose $\exists \: v \in \mu({\cal V} \setminus \mu({\cal V}))$.
\\
As $\mu$ is SC, $\mu({\cal V} \setminus \mu({\cal V})) \cap {\cal V} \subseteq \mu({\cal V})$.
Thus, $v \in \mu({\cal V})$.
Therefore, $v \not\in {\cal V} \setminus \mu({\cal V})$.
\\
But, $\mu({\cal V} \setminus \mu({\cal V})) \subseteq {\cal V} \setminus \mu({\cal V})$.
Thus, $v \not\in \mu({\cal V} \setminus \mu({\cal V}))$, which is impossible.\qed
\end{proof}

\begin{definition}
Let $\langle {\cal F}, {\cal V}, \models \rangle$ be a semantic structure
and $\mid\!\sim$ be a relation on ${\cal P}({\cal F}) \times {\cal F}$.
\\
Then, consider the following conditions: $\forall \: \Gamma \subseteq {\cal F}$,
\begin{description}
\item[$(\mid\!\sim$$4)$] $\C{\mid\!\sim}{\Gamma} =
\T{\lbrace v \in \M{\Gamma} : \forall \: \Delta \subseteq {\cal F},\; \textrm{if}\; v \in \M{\Delta},\;
\textrm{then}\; v \in \M{\C{\mid\!\sim}{\Delta}} \rbrace}$;
\item[$(\mid\!\sim$$5)$] ${\cal V} \setminus \lbrace v \in {\cal V} : \forall \: \Delta \subseteq {\cal F},\; \textrm{if}\; v \in \M{\Delta},\;
\textrm{then}\; v \in \M{\C{\mid\!\sim}{\Delta}} \rbrace \in {\bf D}$.
\end{description}
\end{definition}

\begin{proposition} \label{PIVrepGen}
Let $\langle {\cal F}, {\cal V}, \models \rangle$ be a semantic structure
and $\mid\!\sim$ a relation on ${\cal P}({\cal F}) \times {\cal F}$.
Then:
\begin{description}
\item[$(0)$] $\mid\!\sim$ is a pivotal consequence relation iff
$\mid\!\sim$ satisfies $(\mid\!\sim$$4)$.
\end{description}
Suppose $\langle {\cal F}, {\cal V}, \models \rangle$ satisfies $(A0)$. Then:
\begin{description}
\item[$(1)$] $\mid\!\sim$ is a UC pivotal consequence relation iff
$\mid\!\sim$ satisfies $(\mid\!\sim$$4)$ and $(\mid\!\sim$$5)$.
\end{description}
\end{proposition}

\begin{proof}
{\it Proof of $(0)$.} Direction:~``$\rightarrow$''
\\
There exists a SC choice function $\mu$ from
$\bf D$ to ${\cal P}({\cal V})$ such that $\forall \: \Gamma \subseteq {\cal F}$,
$\C{\mid\!\sim}{\Gamma} = \T{\mu(\M{\Gamma})}$.
\\
Let $f$ be the function from $\bf D$ to $\bf D$ such that
$\forall \: V \in {\bf D}$, we have $f(V) = \M{\T{\mu(V)}}$.
\\
By Lemma~\ref{PIVpivgen}, $\forall \: V \in {\bf D}$, we have $f(V) = \M{\T{\nupp{f}{V}}}$.
\\
Note that $\forall \: \Gamma \subseteq {\cal F}$, $f(\M{\Gamma}) = \M{\T{\mu(\M{\Gamma})}} =
\M{\C{\mid\!\sim}{\Gamma}}$.
\\
We show that $(\mid\!\sim$$4)$ holds.
Let $\Gamma \subseteq {\cal F}$.
\\
Then,
$\C{\mid\!\sim}{\Gamma} =
\T{\mu(\M{\Gamma})} =
\T{\M{\T{\mu(\M{\Gamma})}}} =
\T{f(\M{\Gamma})} =
\T{\M{\T{\nupp{f}{\M{\Gamma}}}}} =
\T{\nupp{f}{\M{\Gamma}}} =
\\
\T{\lbrace v \in \M{\Gamma} : \forall \: W \in {\bf D}$,
if $v \in W$, then $v \in f(W) \rbrace} =
\\
\T{\lbrace v \in \M{\Gamma} : \forall \: \Delta \subseteq {\cal F}$,
if $v \in \M{\Delta}$, then $v \in f(\M{\Delta}) \rbrace} =
\\
\T{\lbrace v \in \M{\Gamma} : \forall \: \Delta \subseteq {\cal F}$,
if $v \in \M{\Delta}$, then $v \in \M{\C{\mid\!\sim}{\Delta}} \rbrace}$.

Direction:~``$\leftarrow$''.
\\
Suppose $\mid\!\sim$ satisfies $(\mid\!\sim$$4)$.
\\
Let $f$ be the function from $\bf D$ to $\bf D$ such that
$\forall \: \Gamma \subseteq {\cal F}$, we have $f(\M{\Gamma}) = \M{\C{\mid\!\sim}{\Gamma}}$.
\\
Note that $f$ is well-defined. Indeed, 
if $\Gamma, \Delta \subseteq {\cal F}$ and $\M{\Gamma} = \M{\Delta}$, then, by $(\mid\!\sim$$4)$,
$\C{\mid\!\sim}{\Gamma} = \C{\mid\!\sim}{\Delta}$.
\\
In addition, by $(\mid\!\sim$$4)$, we clearly have
$\forall \: \Gamma \subseteq {\cal F}$, $\C{\mid\!\sim}{\Gamma} = \T{\nupp{f}{\M{\Gamma}}}$.
\\
And finally, by Lemma~\ref{PIVmufestSCf}, $\nup{f}$ is a SC choice function.
\\ \\
{\it Proof of $(1)$.} Direction: ``$\rightarrow$''.
\\
Verbatim the proof of $(0)$, except that in addition $(A0)$ holds and $\mu$ is UC.
\\
We show that $\mid\!\sim$ satisfies $(\mid\!\sim$$5)$.
As $\mu$ is UC, ${\cal V} \setminus \mu({\cal V}) \in {\bf D}$.
But, by Lemma~\ref{PIVpivgen}, $\mu({\cal V}) = \nupp{f}{{\cal V}} =
\\
\lbrace v \in {\cal V} : \forall \: W \in {\bf D}$, if $v \in W$, then $v \in f(W) \rbrace =
\\
\lbrace v \in {\cal V} : \forall \: \Delta \subseteq {\cal F}$,
if $v \in \M{\Delta}$, then $v \in f(\M{\Delta}) \rbrace =
\\
\lbrace v \in {\cal V} : \forall \: \Delta \subseteq {\cal F}$,
if $v \in \M{\Delta}$, then $v \in \M{\C{\mid\!\sim}{\Delta}} \rbrace$.

Direction: ``$\leftarrow$''.
\\
Verbatim the proof of $(0)$, except that in addition $(A0)$ holds and $\mid\!\sim$ satisfies $(\mid\!\sim$$5)$.
\\
But, because of $(\mid\!\sim$$5)$, ${\cal V} \setminus \nupp{f}{{\cal V}} \in {\bf D}$.
Therefore $\nup{f}$ is UC.
\\
Note that in this direction $(A0)$ is not used.\qed
\end{proof}

%%%%%%%%%%%%%%%%%%%%%%%
\subsection{The discriminative and definability preserving case}\label{PIVdefinablepivotaldiscri}

In the present section, we will characterize certain families of DP pivotal-discriminative consequence relations.
We need an inductive construction introduced in \cite{Bennaim1}:

\begin{notation}
$\mathbb{N}$ denotes the natural numbers: $\lbrace 0, 1, 2, \ldots \rbrace$ and
\\
$\mathbb{N}^{+}$ the positive natural numbers: $\lbrace 1, 2, \ldots \rbrace$.
\end{notation}

\begin{definition}
Let $\cal L$ be a language, $\neg$ a unary connective of $\cal L$, $\cal F$ the set of all wffs of $\cal L$,
$\langle {\cal F}, {\cal V}, \models \rangle$ a semantic structure, $\mid\!\sim$ a relation on
${\cal P}({\cal F}) \times {\cal F}$, and $\Gamma \subseteq {\cal F}$. Then,
$$\monHi{1}{\Gamma} := \lbrace \neg\beta \in {\cal F} :
\beta \in \CC{\vdash}{\Gamma}{\C{\mid\!\sim}{\Gamma}} \setminus \C{\mid\!\sim}{\Gamma} \; \textrm{and}\;
\neg\beta \not\in \CC{\vdash}{\Gamma}{\C{\mid\!\sim}{\Gamma}}\rbrace.$$
Let $i \in \mathbb{N}$ with $i \geq 2$. Then,
$$\monHi{i}{\Gamma} := \lbrace \neg\beta \in {\cal F} :
\left \{ \begin{array}{l}
\beta \in \CCC{\vdash}{\Gamma}{\C{\mid\!\sim}{\Gamma}}
{\monHi{1}{\Gamma}, \ldots, \monHi{i-1}{\Gamma}} \setminus \C{\mid\!\sim}{\Gamma}\; \textrm{and}\\
\neg\beta \not\in \CCC{\vdash}{\Gamma}{\C{\mid\!\sim}{\Gamma}}
{\monHi{1}{\Gamma}, \ldots, \monHi{i-1}{\Gamma}} \end{array} \right . \rbrace.$$
$$\monH{\Gamma} := \bigcup_{i \in \mathbb{N}^+} \monHi{i}{\Gamma}.$$
\end{definition}
We turn to the representation result:

\begin{definition}
Suppose $\cal L$ is a language, $\neg$ a unary connective of $\cal L$,
$\vee$ a binary connective of $\cal L$,
$\cal F$ the set of all wffs of $\cal L$,
$\langle {\cal F}, {\cal V}, \models \rangle$ a semantic structure, and $\mid\!\sim$ a relation on
${\cal P}({\cal F}) \times {\cal F}$.
\\
Then, consider the following conditions: $\forall \: \Gamma, \Delta \subseteq {\cal F}$, $\forall \: \alpha, \beta \in {\cal F}$,
\begin{description}
\item[$(\mid\!\sim$$6)$]
if $\beta \in \CC{\vdash}{\Gamma}{\C{\mid\!\sim}{\Gamma}} \setminus \C{\mid\!\sim}{\Gamma}$ and
$\neg\alpha \in \CCC{\vdash}{\Gamma}{\C{\mid\!\sim}{\Gamma}}{\neg\beta}$, then
$\alpha \not\in \C{\mid\!\sim}{\Gamma}$;
\item[$(\mid\!\sim$$7)$]
if $\alpha \in \CC{\vdash}{\Gamma}{\C{\mid\!\sim}{\Gamma}} \setminus \C{\mid\!\sim}{\Gamma}$ and
$\beta \in \CCC{\vdash}{\Gamma}{\C{\mid\!\sim}{\Gamma}}{\neg\alpha} \setminus \C{\mid\!\sim}{\Gamma}$, then
$\alpha \vee \beta \not\in \C{\mid\!\sim}{\Gamma}$;
\item[$(\mid\!\sim$$8)$]
if $\alpha \in \C{\mid\!\sim}{\Gamma}$, then $\neg\alpha \not\in \CC{\vdash}{\Gamma}{\C{\mid\!\sim}{\Gamma}}$;
\item[$(\mid\!\sim$$9)$] $\C{\mid\!\sim}{\Gamma} \cup  \monH{\Gamma} \subseteq
\CCCC{\vdash}{\Delta}{\C{\mid\!\sim}{\Delta}}{\monH{\Delta}}{\Gamma}$;
\item[$(\mid\!\sim$$10)$]
if $\Gamma$ is consistent, then $\C{\mid\!\sim}{\Gamma}$ is consistent,
$\Gamma \subseteq \C{\mid\!\sim}{\Gamma}$, and
$\C{\vdash}{\C{\mid\!\sim}{\Gamma}} = \C{\mid\!\sim}{\Gamma}$.
\end{description}
\end{definition}
Note that those conditions are purely syntactic
when there is a proof system available for $\vdash$

\begin{proposition} \label{PIVrepArgSyn}
Suppose $\cal L$ is a language, $\neg$ a unary connective of $\cal L$,
$\vee$ and $\wedge$ binary connectives of $\cal L$,
$\cal F$ the set of all wffs of $\cal L$,
$\langle {\cal F}, {\cal V}, \models \rangle$ a semantic structure satisfying $(A3)$ and $(A1)$, and $\mid\!\sim$ a relation on
${\cal P}({\cal F}) \times {\cal F}$. Then:
\begin{description}
\item[$(0)$] $\mid\!\sim$ is a CP DP pivotal-discriminative consequence relation iff
$\mid\!\sim$ satisfies
$(\mid\!\sim$$0)$, $(\mid\!\sim$$6)$, $(\mid\!\sim$$7)$, $(\mid\!\sim$$8)$, $(\mid\!\sim$$9)$, and
$(\mid\!\sim$$10)$.
\end{description}
Suppose $\langle {\cal F}, {\cal V}, \models \rangle$ satisfies $(A2)$. Then:
\begin{description}
\item[$(1)$] $\mid\!\sim$ is a DP pivotal-discriminative consequence relation iff $\mid\!\sim$ satisfies
$(\mid\!\sim$$0)$, $(\mid\!\sim$$6)$, $(\mid\!\sim$$7)$, $(\mid\!\sim$$8)$ and $(\mid\!\sim$$9)$.
\end{description}
\end{proposition}
Before we show Proposition~\ref{PIVrepArgSyn},
we need to introduce Lemmas~\ref{PIV3sim} and \ref{PIVPrf2ConArgSyn} below, taken from \cite{Bennaim1}:

\begin{lemma}\label{PIV3sim} From \cite{Bennaim1}.
\\
Suppose $\cal L$ is a language, $\neg$ a unary connective of $\cal L$,
$\vee$ and $\wedge$ binary connectives of $\cal L$,
$\cal F$ the set of all wffs of $\cal L$,
$\langle {\cal F}, {\cal V}, \models \rangle$ a semantic structure satisfying $(A3)$ and $(A1)$,
and $\mid\!\sim$ a relation on ${\cal P}({\cal F}) \times {\cal F}$ satisfying $(\mid\!\sim$$6)$, $(\mid\!\sim$$7)$, and $(\mid\!\sim$$8)$.
\\
Then, $\forall \: \Gamma \subseteq {\cal F}$,
$\C{\mid\!\sim}{\Gamma} = \Td{\MMM{\Gamma}{\C{\mid\!\sim}{\Gamma}}{\monH{\Gamma}}}$.
\end{lemma}

\begin{lemma} \label{PIVPrf2ConArgSyn} From \cite{Bennaim1}.
\\
Suppose $\cal L$ is a language, $\neg$ a unary connective of $\cal L$,
$\vee$ and $\wedge$ binary connectives of $\cal L$,
$\cal F$ the set of all wffs of $\cal L$,
$\langle {\cal F}, {\cal V}, \models \rangle$ a semantic structure satisfying $(A3)$ and $(A1)$,
${\bf V} \subseteq {\cal P}({\cal V})$,
$\mu$ a DP choice function from $\bf D$ to $\bf V$, $\mid\!\sim$ the
relation on ${\cal P}({\cal F}) \times {\cal F}$ such that
$\forall \: \Gamma \subseteq {\cal F}$, $\C{\mid\!\sim}{\Gamma}
= \Td{\mu(\M{\Gamma})}$, and $\Gamma \subseteq {\cal F}$.
Then:
\begin{description}
\item[$(0)$] $\mid\!\sim$ satisfies $(\mid\!\sim$$6)$, $(\mid\!\sim$$7)$, and $(\mid\!\sim$$8)$;
\item[$(1)$] if $\langle {\cal F}, {\cal V}, \models \rangle$ satisfies $(A2)$ too, then
$\mu(\M{\Gamma}) = \MMM{\Gamma}{\C{\mid\!\sim}{\Gamma}}{\monH{\Gamma}}$;
\item[$(2)$] if $\mu$ is coherency preserving, then
$\mu(\M{\Gamma}) = \MMM{\Gamma}{\C{\mid\!\sim}{\Gamma}}{\monH{\Gamma}}$.
\end{description}
\end{lemma}
We come to the proof of {\bf Proposition~\ref{PIVrepArgSyn}}.

\begin{proof}
{\it Proof of $(0)$.} Direction: ``$\rightarrow$''.
\\
There exists a CP DP SC choice function $\mu$ from $\bf D$ to ${\cal P}({\cal V})$ such that
\\
$\forall \: \Gamma \subseteq {\cal F}$, $\C{\mid\!\sim}{\Gamma} = \Td{\mu(\M{\Gamma})}$.
\\
We will show:
\\
$(0.0)$\quad $\mid\!\sim$ satisfies $(\mid\!\sim$$0)$.
\\
By Lemma~\ref{PIVPrf2ConArgSyn}~$(0)$,
%~$(1)$, $(2)$, and $(3)$,
$\mid\!\sim$ satisfies $(\mid\!\sim$$6)$, $(\mid\!\sim$$7)$, and $(\mid\!\sim$$8)$.
\\
By Lemma~\ref{PIVPrf2ConArgSyn}~$(2)$
%$(10)$
and Strong Coherence of $\mu$, $\mid\!\sim$ satisfies $(\mid\!\sim$$9)$.
\\
We will show:
\\
$(0.1)$\quad $\mid\!\sim$ satisfies $(\mid\!\sim$$10)$.

Direction: ``$\leftarrow$''.
\\
Suppose $\mid\!\sim$ satisfies $(\mid\!\sim$$0)$, $(\mid\!\sim$$6)$, $(\mid\!\sim$$7)$, $(\mid\!\sim$$8)$,
$(\mid\!\sim$$9)$, and $(\mid\!\sim$$10)$.
\\
Then, let $\mu$ be the function from ${\bf D}$ to ${\cal P}({\cal V})$ such that
$\forall \: \Gamma \subseteq {\cal F}$, $\mu(\M{\Gamma}) = \MMM{\Gamma}{\C{\mid\!\sim}{\Gamma}}{\monH{\Gamma}}$.
\\
We will show:
\\
$(0.2)$\quad $\mu$ is well-defined.
\\
Clearly, $\mu$ is a DP choice function.
\\
In addition, as $\mid\!\sim$ satisfies $(\mid\!\sim$$9)$, $\mu$ is strongly coherent.
\\
We will show:
\\
$(0.3)$\quad $\mu$ is CP.
\\
And finally, by Lemma~\ref{PIV3sim},
%~$(7)$,
$\forall \: \Gamma \subseteq {\cal F}$,
$\C{\mid\!\sim}{\Gamma} = \Td{\mu(\M{\Gamma})}$.
\\ \\
{\it Proof of $(0.0)$.} Let $\Gamma, \Delta \subseteq {\cal F}$ and suppose
$\C{\vdash}{\Gamma} = \C{\vdash}{\Delta}$.
Then, $\M{\Gamma} = \M{\Delta}$.
\\
Therefore, $\C{\mid\!\sim}{\Gamma} = \Td{\mu(\M{\Gamma})} = \Td{\mu(\M{\Delta})} = \C{\mid\!\sim}{\Delta}$.
\\ \\
{\it Proof of $(0.1)$.} Let $\Gamma \subseteq {\cal F}$ and suppose $\Gamma$ is consistent.
\\
Then, $\M{\Gamma} \in {\bf D} \cap {\bf C}$. Thus, as $\mu$ is CP, $\mu(\M{\Gamma}) \in {\bf C}$.
Therefore, $\Td{\mu(\M{\Gamma})} = \T{\mu(\M{\Gamma})}$.
\\
Consequently, $\Gamma \subseteq \T{\M{\Gamma}} \subseteq \T{\mu(\M{\Gamma})} = \Td{\mu(\M{\Gamma})} = \C{\mid\!\sim}{\Gamma}$.
\\
In addition, $\M{\C{\mid\!\sim}{\Gamma}} = \M{\Td{\mu(\M{\Gamma})}} = \M{\T{\mu(\M{\Gamma})}}$.
But, $\mu(\M{\Gamma}) \in {\bf C}$. Thus, $\M{\T{\mu(\M{\Gamma})}} \in {\bf C}$.
\\
Consequently, $\C{\mid\!\sim}{\Gamma}$ is consistent.
\\
And finally, $\C{\mid\!\sim}{\Gamma} = \Td{\mu(\M{\Gamma})} =
\T{\mu(\M{\Gamma})} = \T{\M{\T{\mu(\M{\Gamma})}}} = \T{\M{\C{\mid\!\sim}{\Gamma}}} = \C{\vdash}{\C{\mid\!\sim}{\Gamma}}$.
\\ \\
{\it Proof of $(0.2)$.} Let $\Gamma, \Delta \subseteq {\cal F}$ and suppose $\M{\Gamma} = \M{\Delta}$.
\\
Then, $\C{\vdash}{\Gamma} = \C{\vdash}{\Delta}$. Thus, by $(\mid\!\sim$$0)$, $\C{\mid\!\sim}{\Gamma} = \C{\mid\!\sim}{\Delta}$.
\\
Consequently, $\monH{\Gamma} = \monH{\Delta}$. Therefore,
$\MMM{\Gamma}{\C{\mid\!\sim}{\Gamma}}{\monH{\Gamma}} = \MMM{\Delta}{\C{\mid\!\sim}{\Delta}}{\monH{\Delta}}$.
\\ \\
{\it Proof of $(0.3)$.} Suppose $V \in {\bf D} \cap {\bf C}$.
Then, $\exists \: \Gamma \subseteq {\cal F}$, $V = \M{\Gamma}$.
\\
Case~1: $\monHi{1}{\Gamma} \not= \emptyset$.
\\
Thus,
$\exists \: \beta \in {\cal F}$, $\beta \not\in \C{\mid\!\sim}{\Gamma}$ and $\MM{\Gamma}{\C{\mid\!\sim}{\Gamma}} \subseteq \M{\beta}$.
\\
By $(\mid\!\sim$$10)$, $\Gamma \subseteq \C{\mid\!\sim}{\Gamma}$ and $\C{\vdash}{\C{\mid\!\sim}{\Gamma}} = \C{\mid\!\sim}{\Gamma}$.
Thus, $\MM{\Gamma}{\C{\mid\!\sim}{\Gamma}} = \M{\C{\mid\!\sim}{\Gamma}}$.
Thus, $\M{\C{\mid\!\sim}{\Gamma}} \subseteq \M{\beta}$.
\\
Therefore, $\beta \in \T{\M{\C{\mid\!\sim}{\Gamma}}} = \C{\vdash}{\C{\mid\!\sim}{\Gamma}} = \C{\mid\!\sim}{\Gamma}$, which is impossible.
\\
Case~2: $\monHi{1}{\Gamma} = \emptyset$.
\\
Then, $\monH{\Gamma} = \emptyset$.
Thus, $\mu(V) = \mu(\M{\Gamma}) = \MMM{\Gamma}{\C{\mid\!\sim}{\Gamma}}{\monH{\Gamma}} =
\M{\C{\mid\!\sim}{\Gamma}}$.
\\
But, by $(\mid\!\sim$$10)$, $\C{\mid\!\sim}{\Gamma}$ is consistent. Therefore,
$\M{\C{\mid\!\sim}{\Gamma}} \in {\bf C}$.
\\ \\
{\it Proof of $(1)$.} Direction: ``$\rightarrow$''.
\\
Verbatim the proof of $(0)$, except that $\mu$ is no longer CP, whilst $(A2)$ now holds.
\\
Note that, in $(0)$, CP was used only to show $(\mid\!\sim$$9)$ and $(\mid\!\sim$$10)$.
\\
But, $(\mid\!\sim$$10)$ is no longer required to hold and we are going to get
$(\mid\!\sim$$9)$ by another mean.
\\
Indeed, by Lemma~\ref{PIVPrf2ConArgSyn}~$(1)$
%~$(9)$
and Strong Coherence of $\mu$, $(\mid\!\sim$$9)$ holds.

Direction: ``$\leftarrow$''.
\\
Verbatim the proof of $(0)$, except that $(\mid\!\sim$$10)$ does no longer hold, whilst $(A2)$ now holds.
\\
However, in $(0)$, $(\mid\!\sim$$10)$ was used only to show that $\mu$ is CP, which is no longer required.
\\
Note that we do not need to use $(A2)$ in this direction.\qed
\end{proof}

%%%%%%%%%%%%%%%%%%%%%%%
\subsection{The discriminative and not necessarily definability preserving case} \label{PIVpivotaldiscri}

Unlike in Section~\ref{PIVdefinablepivotaldiscri}, the conditions of the present section will not be purely syntactic.
The translation of properties like Strong Coherence in syntactic terms
is blocked because we do no longer have the following useful equality:
$\mu(\M{\Gamma}) = \MMM{\Gamma}{\C{\mid\!\sim}{\Gamma}}{\monH{\Gamma}}$,
which hold when the choice functions under consideration are definability preserving
(but this is not the case here).
Thanks to Lemmas~\ref{PIVmufestSCf} and \ref{PIVpivgen} (stated in Section~\ref{PIVpivotal}),
we will provide a solution with semi-syntactic conditions.

\begin{definition}
Let $\cal L$ be a language, $\neg$ a unary connective of $\cal L$, $\cal F$ the set of all wffs of $\cal L$,
$\langle {\cal F}, {\cal V}, \models \rangle$ a semantic structure, and $\mid\!\sim$ a relation on
${\cal P}({\cal F}) \times {\cal F}$.
\\
Then, consider the following conditions: $\forall \: \Gamma \subseteq {\cal F}$,
\begin{description}
\item[$(\mid\!\sim$$11)$]
$\CCC{\vdash}{\Gamma}{\C{\mid\!\sim}{\Gamma}}{\monH{\Gamma}} =
\T{\lbrace v \in \M{\Gamma} : \forall \: \Delta \subseteq {\cal F},\; \textrm{if}\; v \in \M{\Delta},\;
\textrm{then}\; v \in \MM{\C{\mid\!\sim}{\Delta}}{\monH{\Delta}} \rbrace}$;
\item[$(\mid\!\sim$$12)$] ${\cal V} \setminus \lbrace v \in {\cal V} : \forall \: \Delta \subseteq {\cal F},\; \textrm{if}\; v \in \M{\Delta},\;
\textrm{then}\; v \in \MM{\C{\mid\!\sim}{\Delta}}{\monH{\Delta}} \rbrace \in {\bf D}$.
\end{description}
\end{definition}

\begin{proposition} \label{PIVrepArgSyn2}
Suppose $\cal L$ is a language, $\neg$ a unary connective of $\cal L$,
$\vee$ and $\wedge$ binary connectives of $\cal L$,
$\cal F$ the set of all wffs of $\cal L$,
$\langle {\cal F}, {\cal V}, \models \rangle$ a semantic structure satisfying $(A3)$ and $(A1)$, and $\mid\!\sim$ a relation on
${\cal P}({\cal F}) \times {\cal F}$. Then:
\begin{description}
\item[$(0)$] $\mid\!\sim$ is a CP pivotal-discriminative consequence relation iff 
$\mid\!\sim$ satisfies $(\mid\!\sim$$0)$,
$(\mid\!\sim$$6)$, $(\mid\!\sim$$7)$, $(\mid\!\sim$$8)$, $(\mid\!\sim$$10)$, and $(\mid\!\sim$$11)$.
\end{description}
If $\langle {\cal F}, {\cal V}, \models \rangle$ satisfies $(A0)$ too, then:
\begin{description}
\item[$(1)$] $\mid\!\sim$ is a CP UC pivotal-discriminative consequence relation iff
$\mid\!\sim$ satisfies $(\mid\!\sim$$0)$,
$(\mid\!\sim$$6)$, $(\mid\!\sim$$7)$, $(\mid\!\sim$$8)$, $(\mid\!\sim$$10)$, $(\mid\!\sim$$11)$, and $(\mid\!\sim$$12)$.
\end{description}
If $\langle {\cal F}, {\cal V}, \models \rangle$ satisfies $(A2)$ too, then:
\begin{description}
\item[$(2)$] $\mid\!\sim$ is a pivotal-discriminative consequence relation iff
$\mid\!\sim$ satisfies
$(\mid\!\sim$$0)$,
$(\mid\!\sim$$6)$, $(\mid\!\sim$$7)$, $(\mid\!\sim$$8)$, and $(\mid\!\sim$$11)$.
\end{description}
If $\langle {\cal F}, {\cal V}, \models \rangle$ satisfies $(A0)$ and $(A2)$ too, then:
\begin{description}
\item[$(3)$] $\mid\!\sim$ is a UC pivotal-discriminative consequence relation iff
$\mid\!\sim$ satisfies
$(\mid\!\sim$$0)$,
$(\mid\!\sim$$6)$, $(\mid\!\sim$$7)$, $(\mid\!\sim$$8)$, $(\mid\!\sim$$11)$, and $(\mid\!\sim$$12)$.
\end{description}
\end{proposition}

\begin{proof}
{\it Proof of $(2)$.} Direction: ``$\rightarrow$''.
\\
There exists a SC choice function $\mu$ from $\bf D$ to ${\cal P}({\cal V})$ such that
$\forall \: \Gamma \subseteq {\cal F}$, $\C{\mid\!\sim}{\Gamma} = \Td{\mu(\M{\Gamma})}$.
\\
Then, $\mid\!\sim$ satisfies obviously $(\mid\!\sim$$0)$.
\\
Let $f$ be the function from ${\bf D}$ to ${\bf D}$ such that
$\forall \: V \in {\bf D}$, $f(V) = \M{\T{\mu(V)}}$.
\\
Then, by Lemma~\ref{PIVpivgen}, $\forall \: V \in {\bf D}$, $f(V) = \M{\T{\nupp{f}{V}}}$.
\\
Moreover, $\forall \: \Gamma \subseteq {\cal F}$,
$f(\M{\Gamma}) = \M{\T{\mu(\M{\Gamma})}} \subseteq \M{\T{\M{\Gamma}}} = \M{\Gamma}$.
\\
Therefore, $f$ is a choice function.
\\
Obviously, $f$ is DP.
\\
In addition, $\forall \: \Gamma \subseteq {\cal F}$,
$\C{\mid\!\sim}{\Gamma} = \Td{\mu(\M{\Gamma})} = \Td{\M{\T{\mu(\M{\Gamma})}}} = \Td{f(\M{\Gamma})}$.
\\
Consequently, by Lemma~\ref{PIVPrf2ConArgSyn}~$(0)$,
%$(1)$, $(2)$, and $(3)$,
$\mid\!\sim$ satisfies
$(\mid\!\sim$$6)$, $(\mid\!\sim$$7)$, and $(\mid\!\sim$$8)$. 
\\
In addition, by Lemma~\ref{PIVPrf2ConArgSyn}~$(1)$,
%~$(9)$,
$\forall \: \Gamma \subseteq {\cal F}$, $f(\M{\Gamma}) =
\MMM{\Gamma}{\C{\mid\!\sim}{\Gamma}}{\monH{\Gamma}}$.
\\
We show that $\mid\!\sim$ satisfies $(\mid\!\sim$$11)$. Let $\Gamma \subseteq {\cal F}$.
\\
Then, $\CCC{\vdash}{\Gamma}{\C{\mid\!\sim}{\Gamma}}{\monH{\Gamma}} =
\T{\MMM{\Gamma}{\C{\mid\!\sim}{\Gamma}}{\monH{\Gamma}}} =
\T{f(\M{\Gamma})} = \T{\M{\T{\nupp{f}{\M{\Gamma}}}}} = \T{\nupp{f}{\M{\Gamma}}} =$
\\
$\T{\lbrace v \in \M{\Gamma} : \forall \: W \in {\bf D}$,
if $v \in W$, then $v \in f(W) \rbrace} =$
\\
$\T{\lbrace v \in \M{\Gamma} : \forall \: \Delta \subseteq {\cal F}$,
if $v \in \M{\Delta}$, then $v \in f(\M{\Delta}) \rbrace} =$
\\
$\T{\lbrace v \in \M{\Gamma} : \forall \: \Delta \subseteq {\cal F}$,
if $v \in \M{\Delta}$, then $v \in \MMM{\Delta}{\C{\mid\!\sim}{\Delta}}{\monH{\Delta}} \rbrace} =$
\\
$\T{\lbrace v \in \M{\Gamma} : \forall \: \Delta \subseteq {\cal F},\; \textrm{if}\; v \in \M{\Delta},\;\textrm{then}\; v \in \MM{\C{\mid\!\sim}{\Delta}}{\monH{\Delta}} \rbrace}$. 

Direction: ``$\leftarrow$''.
\\
Suppose $(\mid\!\sim$$0)$, $(\mid\!\sim$$6)$, $(\mid\!\sim$$7)$, $(\mid\!\sim$$8)$, and $(\mid\!\sim$$11)$ hold.
\\
Let $f$ be the function from $\bf D$ to $\bf D$ such that
$\forall \: \Gamma \subseteq {\cal F}$, $f(\M{\Gamma}) =
\MMM{\Gamma}{\C{\mid\!\sim}{\Gamma}}{\monH{\Gamma}}$.
\\
By $(\mid\!\sim$$0)$, $f$ is well-defined.
\\
By Lemma~\ref{PIV3sim},
%~$(7)$,
$\forall \: \Gamma \subseteq {\cal F}$, $\C{\mid\!\sim}{\Gamma} = \Td{\MMM{\Gamma}{\C{\mid\!\sim}{\Gamma}}{\monH{\Gamma}}} = \Td{f(\M{\Gamma})}$.
\\
By $(\mid\!\sim$$11)$, $\forall \: \Gamma \subseteq {\cal F}$, $f(\M{\Gamma}) = \M{\T{\nupp{f}{\M{\Gamma}}}}$.
\\
Therefore, $\forall \: \Gamma \subseteq {\cal F}$, $\C{\mid\!\sim}{\Gamma} = \Td{f(\M{\Gamma})}
= \Td{\M{\T{\nupp{f}{\M{\Gamma}}}}} = \Td{\nupp{f}{\M{\Gamma}}}$.
\\
But, by Lemma~\ref{PIVmufestSCf}, $\nup{f}$ is a SC choice function.
\\ \\
{\it Proof of $(3)$}. Direction: ``$\rightarrow$''.
\\
Verbatim the proof of $(2)$, except that in addition $(A0)$ holds and $\mu$ is UC.
\\
We show that $(\mid\!\sim$$12)$ holds.
As $\mu$ is UC, ${\cal V} \setminus \mu({\cal V}) \in {\bf D}$.
But, by Lemma~\ref{PIVpivgen}~$(1)$, $\mu({\cal V}) = \nupp{f}{{\cal V}} =
\\
\lbrace v \in {\cal V} : \forall \: W \in {\bf D}$, if $v \in W$, then $v \in f(W) \rbrace =
\\
\lbrace v \in {\cal V} : \forall \: \Delta \subseteq {\cal F}$,
if $v \in \M{\Delta}$, then $v \in f(\M{\Delta}) \rbrace =
\\
\lbrace v \in {\cal V} : \forall \: \Delta \subseteq {\cal F},\; \textrm{if}\; v \in \M{\Delta},\;
\textrm{then}\; v \in \MMM{\Delta}{\C{\mid\!\sim}{\Delta}}{\monH{\Delta}} \rbrace =
\\
\lbrace v \in {\cal V} : \forall \: \Delta \subseteq {\cal F},\; \textrm{if}\; v \in \M{\Delta},\;
\textrm{then}\; v \in \MM{\C{\mid\!\sim}{\Delta}}{\monH{\Delta}} \rbrace$.

Direction: ``$\leftarrow$''.
\\
Verbatim the proof of $(2)$, except that in addition $(A0)$ holds and $\mid\!\sim$ satisfies $(\mid\!\sim$$12)$.
\\
But, because of $(\mid\!\sim$$12)$, ${\cal V} \setminus \nupp{f}{{\cal V}} \in {\bf D}$.
Therefore $\nup{f}$ is UC.
\\
Note that $(A0)$ is not used in this direction.
\hfill \\ \\
{\it Proof of $(0)$.} Direction: ``$\rightarrow$''.
\\
Verbatim the proof of $(2)$, except that $(A2)$ does no longer hold, whilst $\mu$ is now CP.
\\
Note that $(A2)$ was used, in $(2)$, only to apply Lemma~\ref{PIVPrf2ConArgSyn}~$(1)$ to get
$\forall \: \Gamma \subseteq {\cal F}$, $f(\M{\Gamma}) =
\MMM{\Gamma}{\C{\mid\!\sim}{\Gamma}}{\monH{\Gamma}}$.
\\
But, we will get this equality by another mean.
\\
Indeed, if $V \in {\bf D} \cap {\bf C}$, then, as $\mu$ is CP, $\mu(V) \in {\bf C}$, thus
$\M{\T{\mu(V)}} \in {\bf C}$, thus $f(V) \in {\bf C}$.
\\
Therefore $f$ is CP.
\\
Consequently, by Lemma~\ref{PIVPrf2ConArgSyn}~$(2)$, we get
$\forall \: \Gamma \subseteq {\cal F}$, $f(\M{\Gamma}) =
\MMM{\Gamma}{\C{\mid\!\sim}{\Gamma}}{\monH{\Gamma}}$.
\\
In addition, by verbatim the proof of $(0.1)$ of Proposition~\ref{PIVrepArgSyn}, $\mid\!\sim$ satisfies $(\mid\!\sim$$10)$.

Direction: ``$\leftarrow$''.
\\
Verbatim the proof of $(2)$, except that $(A2)$ does no longer hold, whilst $\mid\!\sim$ satisfies now
$(\mid\!\sim$$10)$.
\\
But, in this direction, $(A2)$ was not used in $(2)$.
\\
It remains to show that $\nup{f}$ is CP.
\\
By verbatim the proof of $(0.3)$ of Proposition~\ref{PIVrepArgSyn}, we get that $f$ is CP.
\\
Let $V \in {\bf D} \cap {\bf C}$. Then, $f(V) \in {\bf C}$. Thus, $\M{\T{\nupp{f}{V}}} \in {\bf C}$.
Thus, $\nupp{f}{V} \in {\bf C}$ and we are done.
\\ \\
{\it Proof of $(1)$.} Direction: ``$\rightarrow$''.
\\
Verbatim the proof of $(2)$, except that $(A2)$ does no longer hold, whilst $(A0)$ now holds
and $\mu$ is now UC and CP.
\\
Note that $(A2)$ was used, in $(2)$, only to apply Lemma~\ref{PIVPrf2ConArgSyn}~$(1)$ to get
$\forall \: \Gamma \subseteq {\cal F}$, $f(\M{\Gamma}) =
\MMM{\Gamma}{\C{\mid\!\sim}{\Gamma}}{\monH{\Gamma}}$.
\\
But, by verbatim the proof of $(0)$, we get anyway
$\forall \: \Gamma \subseteq {\cal F}$, $f(\M{\Gamma}) =
\MMM{\Gamma}{\C{\mid\!\sim}{\Gamma}}{\monH{\Gamma}}$.
\\
In addition, by verbatim the proof of $(0.1)$ of Proposition~\ref{PIVrepArgSyn}, $\mid\!\sim$ satisfies $(\mid\!\sim$$10)$.
\\
And, by verbatim the proof of $(3)$, $\mid\!\sim$ satisfies $(\mid\!\sim$$12)$.

Direction: ``$\leftarrow$''.
\\
Verbatim the proof of $(2)$, except that $(A2)$ does no longer hold, whilst
$(A0)$ now holds and $\mid\!\sim$ satisfies now $(\mid\!\sim$$10)$ and $(\mid\!\sim$$12)$.
\\
But, in this direction, $(A2)$ was not used in $(2)$.
\\
In addition, by verbatim the proof of $(0)$, $\nup{f}$ is CP.
\\
And, because of $(\mid\!\sim$$12)$, ${\cal V} \setminus \nupp{f}{{\cal V}} \in {\bf D}$.
Therefore $\nup{f}$ is UC.
\\
Note that $(A0)$ is not used in this direction.\qed
\end{proof}

%%%%%%%%%%%%%%%%%%%%%%%
\section{Nonexistence of normal characterizations} \label{PIVsectionNoCharac}

%%%%%%%%%
\subsection{Definition}

Let $\cal F$ be a set, $\cal R$ a set of relations on ${\cal P}({\cal F}) \times {\cal F}$,
and $\mid\!\sim$ a relation on ${\cal P}({\cal F}) \times {\cal F}$.
\\
Approximatively, a characterization of $\cal R$ will be called ``normal'' iff
it contains only conditions which are universally quantified and ``apply'' $\mid\!\sim$ at most $|{\cal F}|$ times.
More formally,

\begin{definition}\label{PIVnormalchar}
Let $\cal F$ be a set and ${\cal R}$ a set of relations on ${\cal P}({\cal F}) \times {\cal F}$.
\\
We say that that $\cal C$ is a {\it normal characterization} of $\cal R$ iff
${\cal C} = \langle \lambda, \Phi \rangle$, where
$\lambda \leq |{\cal F}|$ is a (finite or infinite) cardinal and
$\Phi$ is a relation on ${\cal P}({\cal F})^{2\lambda}$ such that for every relation $\mid\!\sim$
on ${\cal P}({\cal F}) \times {\cal F}$,
$$\mid\!\sim\: \in {\cal R} \;\:\textrm{iff}\;\: \forall \: \Gamma_1, \ldots, \Gamma_\lambda \subseteq {\cal F},\;
(\Gamma_1, \ldots, \Gamma_\lambda, \C{\mid\!\sim}{\Gamma_1}, \ldots, \C{\mid\!\sim}{\Gamma_\lambda})
\in \Phi.$$
\end{definition}
Now, suppose there is no normal characterization of $\cal R$.
Here are examples (i.e. $(C1)$, $(C2)$, and $(C3)$ below) that will give the reader (we hope) a good idea
which conditions cannot characterize $\cal R$. This will thus make clearer the range
of our impossibility result (Proposition~\ref{PIVnoCharac} below). To begin, consider the following condition:
\begin{description}
\item[$(C1)$] $\forall \: \Gamma, \Delta \in {\bf F} \subseteq {\cal P}({\cal F})$,
$\C{\mid\!\sim}{\Gamma \cup \C{\mid\!\sim}{\Delta}} = \emptyset$.
\end{description}
Then, $(C1)$ cannot characterize $\cal R$.
Indeed, suppose the contrary, i.e.
\\
suppose $\mid\!\sim \: \in {\cal R}$ iff
$\forall \: \Gamma, \Delta \in {\bf F}$, $\C{\mid\!\sim}{\Gamma \cup \C{\mid\!\sim}{\Delta}} = \emptyset$.
\\
Then, take $\lambda = 3$ and the relation $\Phi$ such that
$(\Gamma_1, \Gamma_2, \Gamma_3, \Gamma_4, \Gamma_5, \Gamma_6) \in \Phi$ iff
\\
$(\Gamma_1, \Gamma_2 \in {\bf F}$ and
$\Gamma_3 = \Gamma_1 \cup \Gamma_5)$ entails $\Gamma_6 = \emptyset$.
\\
Then, $\langle 3, \Phi \rangle$ is a normal characterization of $\cal R$. 
We give the easy proof of this, so that the reader can check that a convenient relation $\Phi$
can be found quickly for all simple conditions like $(C1)$.
\begin{proof}
Direction: ``$\rightarrow$''.
\\
Suppose $\mid\!\sim \: \in {\cal R}$.
\\
Then, $\forall \: \Gamma, \Delta \in {\bf F}$,
$\C{\mid\!\sim}{\Gamma \cup \C{\mid\!\sim}{\Delta}} = \emptyset$.
\\
Let $\Gamma_1, \Gamma_2, \Gamma_3 \subseteq {\cal F}$.
\\
We show $(\Gamma_1, \Gamma_2, \Gamma_3, \C{\mid\!\sim}{\Gamma_1}, \C{\mid\!\sim}{\Gamma_2},
\C{\mid\!\sim}{\Gamma_3})
\in \Phi$.
\\
Suppose $\Gamma_1, \Gamma_2 \in {\bf F}$ and $\Gamma_3 = \Gamma_1 \cup \C{\mid\!\sim}{\Gamma_2}$.
\\
Then, as $\Gamma_1, \Gamma_2 \in {\bf F}$, we get $\C{\mid\!\sim}{\Gamma_1 \cup \C{\mid\!\sim}{\Gamma_2}} = \emptyset$.
\\
But, $\C{\mid\!\sim}{\Gamma_1 \cup \C{\mid\!\sim}{\Gamma_2}} = \C{\mid\!\sim}{\Gamma_3}$.
Therefore, $\C{\mid\!\sim}{\Gamma_3} = \emptyset$.

Direction: ``$\leftarrow$''.
\\
Suppose $\forall \: \Gamma_1, \Gamma_2, \Gamma_3 \subseteq {\cal F},\;
(\Gamma_1, \Gamma_2, \Gamma_3, \C{\mid\!\sim}{\Gamma_1}, \C{\mid\!\sim}{\Gamma_2},
\C{\mid\!\sim}{\Gamma_3})
\in \Phi$.
\\
We show $\mid\!\sim \: \in {\cal R}$.
Let $\Gamma, \Delta \in {\bf F}$.
\\
Take $\Gamma_1 = \Gamma$,
$\Gamma_2 = \Delta$,
$\Gamma_3 = \Gamma_1 \cup \C{\mid\!\sim}{\Gamma_2}$.
\\
Then, we have $\Gamma_1 \in {\bf F}$, $\Gamma_2 \in {\bf F}$,
and $\Gamma_3 = \Gamma_1 \cup \C{\mid\!\sim}{\Gamma_2}$.
\\
But, $(\Gamma_1, \Gamma_2, \Gamma_3, \C{\mid\!\sim}{\Gamma_1}, \C{\mid\!\sim}{\Gamma_2},
\C{\mid\!\sim}{\Gamma_3}) \in \Phi$.
\\
Therefore, by definition of $\Phi$, $\C{\mid\!\sim}{\Gamma_3} = \emptyset$.
\\
But, $\C{\mid\!\sim}{\Gamma_3} = \C{\mid\!\sim}{\Gamma_1 \cup \C{\mid\!\sim}{\Gamma_2}} =
\C{\mid\!\sim}{\Gamma \cup \C{\mid\!\sim}{\Delta}}$.\qed
\end{proof}
But actually, we are not limited to simple operations (like e.g. $\cup$, $\cap$, $\setminus$).
More complex conditions than $(C1)$ are also excluded. For instance, let $f$ be any function
from ${\cal P}({\cal F})$ to ${\cal P}({\cal F})$ and consider the following condition:
\begin{description}
\item[$(C2)$] $\forall \: \Gamma, \Delta \in {\bf F}$,
$\C{\mid\!\sim}{f(\Gamma) \cup \C{\mid\!\sim}{\Delta}} = \emptyset$.
\end{description}
Then, $(C2)$ cannot characterize $\cal R$. Indeed, suppose it characterizes $\cal R$.
\\
Then, take the relation $\Phi$ such that
$(\Gamma_1, \Gamma_2, \Gamma_3, \Gamma_4, \Gamma_5, \Gamma_6) \in \Phi$ iff
\\
$(\Gamma_1, \Gamma_2 \in {\bf F}$ and
$\Gamma_3 = f(\Gamma_1) \cup \Gamma_5)$ entails $\Gamma_6 = \emptyset$.
\\
It can be checked that $\langle 3, \Phi \rangle$ is a normal characterization of $\cal R$.
We leave the easy proof to the reader.

We can even go further combining universal (not existential) quantifiers and functions like~$f$.
For instance, let ${\cal G}$ be a set of functions from ${\cal P}({\cal F})$ to ${\cal P}({\cal F})$ and
consider the following condition:
\begin{description}
\item[$(C3)$] $\forall \: \Gamma, \Delta \in {\bf F}$, $\forall \: f \in {\cal G}$,
$\C{\mid\!\sim}{f(\Gamma) \cup \C{\mid\!\sim}{\Delta}} = \emptyset$.
\end{description}
Then, $(C3)$ cannot characterize $\cal R$. Indeed, suppose it characterizes $\cal R$.
\\
Then, take the relation $\Phi$ such that
$(\Gamma_1, \Gamma_2, \Gamma_3, \Gamma_4, \Gamma_5, \Gamma_6) \in \Phi$ iff
\\
$\forall \: f \in {\cal G}$, if $(\Gamma_1, \Gamma_2 \in {\bf F}$ and
$\Gamma_3 = f(\Gamma_1) \cup \Gamma_5)$, then $\Gamma_6 = \emptyset$.
\\
It can be checked that $\langle 3, \Phi \rangle$ is a normal characterization of $\cal R$.
The easy proof is left to the reader.

Finally, a good example of a condition which is not excluded is $(\mid\!\sim$$4)$.
We have seen in Proposition~\ref{PIVrepGen} that it characterizes the family of all
pivotal consequence relations.

%%%%%%%%%
\subsection{Impossibility results}

In the present section, we will show, in an infinite classical framework,
that there is no normal characterization for
the family of all pivotal consequence relations (in other words,
$(\mid\!\sim$$4)$ cannot be replaced by a simpler condition in Proposition~\ref{PIVrepGen}).
In the same vein, in Proposition~5.2.15 of \cite{Schlechta5}, K. Schlechta showed
that there does not exist a normal characterization for the family of all
preferential consequence relations.

Note that he used the word ``normal'' in a more restrictive sense (see Section~1.6.2.1 of \cite{Schlechta5}).
Approximatively, a characterization of $\cal R$ is called normal by Schlechta iff it contains only conditions like $(C1)$,
i.e. conditions which are universally quantified, ``apply'' $\mid\!\sim$ at most $|{\cal F}|$ times, and use only
elementary operations like e.g. $\cup$, $\cap$, $\setminus$
(complex structures, functions, etc are not allowed).
We have been inspired by the techniques of Schlechta.
We will need Lemma 5.2.14 of \cite{Schlechta5}:

\begin{lemma} \label{PIVideal}
From \cite{Schlechta5}.
\\
Suppose $\cal A$ is infinite and $\langle {\cal F}_c, {\cal V}, \models \rangle$
is a classical propositional semantic structure.
\\
Let ${\bf V} \subseteq \lbrace V \subseteq {\cal V} : |V| \leq |{\cal A}| \rbrace$
satisfying the two following conditions:
\\
first, if $V \in {\bf V}$ and $W \subseteq V$, then $W \in {\bf V}$;
\\
and second, $\forall \: V, W \in {\bf V}$, if $|V \cup W| \leq |{\cal A}|$, then $V \cup W \in {\bf V}$.
\\
Then, $\forall \: \Gamma \subseteq {\cal F}_c$, $\exists \: V_\Gamma \in {\bf V}$,
\begin{description}
\item[$(0)$] $\T{\bigcap_{V \in {\bf V}} \M{\T{\M{\Gamma} \setminus V}}} = \T{\M{\Gamma} \setminus V_\Gamma}$;
\item[$(1)$] $\forall \: V \in {\bf V}$, $\T{\M{\Gamma} \setminus V} \subseteq \T{\M{\Gamma} \setminus V_\Gamma}$.
\end{description}
\end{lemma}
Recall that $\cal A$ and ${\cal F}_c$ have been introduced in Section~\ref{PIVfourframework}.
Note that the subscript in $V_\Gamma$ is written just to keep in mind that $V_\Gamma$ depends on $\Gamma$.

\begin{proposition} \label{PIVnoCharac}
Suppose $\cal A$ is infinite and $\langle {\cal F}_c, {\cal V}, \models \rangle$
is a classical propositional semantic structure.
\\
Then, there doesn't exist a normal characterization for the family of all pivotal consequence relations.
\end{proposition}

\begin{proof}
Suppose the contrary, i.e. suppose there exist a cardinal $\lambda \leq |{{\cal F}_c}|$
and a relation $\Phi$ over ${\cal P}({{\cal F}_c})^{2\lambda}$ such that
for every relation $\mid\!\sim$ on ${\cal P}({\cal F}_c) \times {\cal F}_c$,
$\mid\!\sim$ is a pivotal consequence relation iff
$\forall \: \Gamma_1, \ldots, \Gamma_\lambda \subseteq {\cal F}_c$,
$(\Gamma_1, \ldots, \Gamma_\lambda, \C{\mid\!\sim}{\Gamma_1}, \ldots, \C{\mid\!\sim}{\Gamma_\lambda})
\in \Phi$.
Then, define:
\\
$
{\bf V} := \lbrace V \subseteq {\cal V} : |V| \leq |{\cal A}| \rbrace.
$
\\
In addition, let $\mid\!\sim$ be the relation on ${\cal P}({\cal F}_c) \times {\cal F}_c$ such that
$\forall \: \Gamma \subseteq {\cal F}_c$,
\\
$
\C{\mid\!\sim}{\Gamma} = T ( \bigcap_{V \in {\bf V}} \M{\T{\M{\Gamma} \setminus V}}).
$
\\
We will show:
\\
$(0)$\quad $\forall \: V \subseteq {\cal V}$, if $|V| \leq |{\cal A}|$, then $\T{{\cal V}} = \T{{\cal V} \setminus V}$;
\\
$(1)$\quad $\exists \: \Gamma_1, \ldots, \Gamma_\lambda \subseteq{\cal F}_c$ such that
$(\Gamma_1, \ldots, \Gamma_\lambda$, $\C{\mid\!\sim}{\Gamma_1}, \ldots, \C{\mid\!\sim}{\Gamma_\lambda})
\not\in \Phi$.
\\
Now, by lemma \ref{PIVideal}, we get:
\\
$(2)$\quad $\forall \: \Gamma \subseteq {\cal F}_c$, $\exists \: V_\Gamma \in {\bf V}$,
$\C{\mid\!\sim}{\Gamma} = \T{\M{\Gamma} \setminus V_\Gamma}$ and
$\forall \: V \in {\bf V}$, $\T{\M{\Gamma} \setminus V} \subseteq \T{\M{\Gamma} \setminus V_\Gamma}$.
\\
Then, define:
\\
$
{\cal X} := \bigcup_{\Gamma \in \lbrace \Gamma_1, \ldots, \Gamma_\lambda \rbrace} V_\Gamma.
$
\\
Then, we will show:
\\
$(3)$\quad $\forall \: \Gamma \in \lbrace \Gamma_1, \ldots, \Gamma_\lambda \rbrace$,
$\C{\mid\!\sim}{\Gamma} = \T{\M{\Gamma} \setminus {\cal X}}$.
\\
Let $\mu$ be the function from $\bf D$ to ${\cal P}({\cal V})$ such that
$\forall \: V \in {\bf D}$,
$\mu(V) = V \setminus {\cal X}$.
\\
We will show:
\\
$(4)$\quad $\mu$ is a SC choice function.
\\
Let $\mid\!\sim'$ be the pivotal consequence relation defined by $\mu$.
\\
We will show the following, which entails a contradiction:
\\
$(5)$\quad $\mid\!\sim'$ is not a pivotal consequence relation.
\\ \\
{\it Proof of $(0)$.} Let $V \subseteq {\cal V}$ and suppose $|V| \leq |{\cal A}|$.
\\
Obviously, $\T{{\cal V}} \subseteq \T{{\cal V} \setminus V}$.
\\
We show $\T{{\cal V} \setminus V} \subseteq \T{{\cal V}}$.
\\
Suppose the contrary, i.e. suppose $\exists \: \alpha \in \T{{\cal V} \setminus V}$, $\alpha \not\in \T{{\cal V}}$.
\\
Then, $\exists \: v \in {\cal V}$, $v \not\in \M{\alpha}$.
\\
Now, define:
\\
$
W := \lbrace w \in {\cal V} : \textrm{for all atom $q$ occurring in $\alpha$, $w(q) = v(q)$} \rbrace.
$
\\
Then, $\forall \: w \in W$, we have $w(\alpha) = v(\alpha)$ and thus $w \not\in \M{\alpha}$.
\\
As the number of atoms occurring in $\alpha$ is finite
and $\cal A$ is infinite, we get $|W| = 2^{|{\cal A}|}$.
\\
Therefore, $|V| \leq |{\cal A}| < |W|$.
Thus, $\exists \: w \in W \setminus V \subseteq {\cal V} \setminus V$.
\\
Thus, ${\cal V} \setminus V \not\subseteq \M{\alpha}$.
Therefore, $\alpha \not\in \T{{\cal V} \setminus V}$, which is impossible.
\\ \\
{\it Proof of $(1)$.} It suffices to show that $\mid\!\sim$ is not a pivotal consequence relation.
\\
Suppose the contrary, i.e. suppose there exists a SC choice function $\mu$ from $\bf D$ to ${\cal P}({\cal V})$
such that
\\
$\forall \: \Gamma \subseteq {\cal F}_c$, $\C{\mid\!\sim}{\Gamma} = \T{\mu(\M{\Gamma})}$.
\\
As $\cal A$ is infinite, $\exists \: p \in {\cal A}$. We show that all cases are impossible.
\\
Case 1: $\exists \: v \in \mu({\cal V})$, $v \not\in \M{p}$.
\\
Let $\Gamma = \T{v}$. Then, $\M{\Gamma} = \lbrace v \rbrace$.
\\
By SC of $\mu$, we have $\mu(\M{\Gamma}) = \mu(\M{\Gamma}) \cap {\cal V} \subseteq \mu({\cal V})$.
Thus, $\mu(\M{\Gamma}) \subseteq \mu({\cal V}) \cap \M{\Gamma}$.
\\
On the other hand, again by SC, $\mu({\cal V}) \cap \M{\Gamma} \subseteq \mu(\M{\Gamma})$.
Consequently, $\mu({\cal V}) \cap \M{\Gamma} = \mu(\M{\Gamma})$.
\\
Therefore, $\C{\mid\!\sim}{\Gamma} = \T{\mu(\M{\Gamma})} =
\T{\mu({\cal V}) \cap \M{\Gamma}} = \T{\mu({\cal V}) \cap \lbrace v \rbrace} = \T{v}$.
\\
But, $p \not\in \T{v}$. Thus, $p \not\in \C{\mid\!\sim}{\Gamma}$.
\\
However, $\M{\Gamma} \in {\bf V}$. Therefore,
$\bigcap_{V \in {\bf V}} \M{\T{\M{\Gamma} \setminus V}} \subseteq \M{\T{\M{\Gamma} \setminus \M{\Gamma}}}
= \M{\T{\emptyset}}
= \M{{\cal F}_c} = \emptyset$.
\\
Therefore, by definition of $\mid\!\sim$, we have $\C{\mid\!\sim}{\Gamma} = \T{\emptyset} = {\cal F}_c$.
\\
Thus, $p \in \C{\mid\!\sim}{\Gamma}$, which is impossible.
\\
Case 2: $\mu({\cal V}) \subseteq \M{p}$.
\\
Then, by $(0)$, $\C{\mid\!\sim}{\emptyset} = \T{\bigcap_{V \in {\bf V}} \M{\T{{\cal V} \setminus V}}} =
\T{\bigcap_{V \in {\bf V}} \M{\T{{\cal V}}}} = \T{\M{\T{{\cal V}}}} = \T{{\cal V}}$.
\\
But, ${\cal V} \not\subseteq \M{p}$. Thus, $p \not\in \T{{\cal V}} = \C{\mid\!\sim}{\emptyset}$.
\\
On the other hand, $\C{\mid\!\sim}{\emptyset} =
\T{\mu(\M{\emptyset})} = \T{\mu({\cal V})}$.
\\
But, $\mu({\cal V}) \subseteq \M{p}$. Thus, $p \in \T{\mu({\cal V})} = \C{\mid\!\sim}{\emptyset}$,
which is impossible.
\\ \\
{\it Proof of $(3)$.} Let $\Gamma \in \lbrace \Gamma_1, \ldots, \Gamma_\lambda \rbrace$.
Direction: ``$\subseteq$''.
\\
We have $V_\Gamma \subseteq {\cal X}$.
Thus, $\M{\Gamma} \setminus {\cal X} \subseteq \M{\Gamma} \setminus V_\Gamma$.
\\
Therefore, by $(2)$, $\C{\mid\!\sim}{\Gamma} = \T{\M{\Gamma} \setminus V_\Gamma} \subseteq \T{\M{\Gamma} \setminus {\cal X}}$.

Direction: ``$\supseteq$''.
\\
As $\cal A$ is infinite, $|{\cal A}| = |{{\cal F}_c}|$.
Therefore, $\lambda \leq |{\cal A}|$.
Thus, $|{\cal X}| \leq |{\cal A}|^2 = |{\cal A}|$.
\\
Thus, ${\cal X} \in {\bf V}$. Thus, by $(2)$, $\T{\M{\Gamma} \setminus {\cal X}} \subseteq \T{\M{\Gamma} \setminus V_\Gamma} = \C{\mid\!\sim}{\Gamma}$.
\\ \\
{\it Proof of $(4)$.}
$\mu$ is clearly a choice function. We show that $\mu$ satisfies SC.
Let $V, W \subseteq {\cal V}$.
\\
Then, $\mu(W) \cap V = (W \setminus {\cal X}) \cap V
= (W \cap V) \setminus {\cal X} \subseteq V \setminus {\cal X} = \mu(V)$.
\\ \\
{\it Proof of $(5)$.}
By $(3)$,
$\forall \: \Gamma \in \lbrace \Gamma_1, \ldots, \Gamma_\lambda \rbrace$,
$\C{\mid\!\sim'}{\Gamma} =
\T{\mu(\M{\Gamma})} = \T{\M{\Gamma} \setminus {\cal X}} = \C{\mid\!\sim}{\Gamma}$.
\\
But, $(\Gamma_1, \ldots, \Gamma_\lambda$, $\C{\mid\!\sim}{\Gamma_1}, \ldots, \C{\mid\!\sim}{\Gamma_\lambda}) \not\in \Phi$.
Therefore, $(\Gamma_1, \ldots, \Gamma_\lambda$, $\C{\mid\!\sim'}{\Gamma_1}, \ldots, \C{\mid\!\sim'}{\Gamma_\lambda}) \not\in \Phi$.
\\
Consequently, as $\langle \lambda, \Phi \rangle$ is a normal characterization,
$\mid\!\sim'$ is not a pivotal consequence relation.\qed
\end{proof}

%%%%%%%%%%%%%%%%%%%%%%%
\section{A link with $X$-logics} \label{PIVsectioncodefisclosed}

In this section, we investigate a link between pivotal consequence relations
and pertinence consequence relations (alias $X$-logics) which were
first introduced by Forget, Risch, and Siegel \cite{ForgetRischSiegel1}.
Suppose some formulas are considered to be the pertinent ones in the absolute sense
and collect them in a set~${\cal E}$.
Then, it is natural to conclude a formula $\alpha$ from a set of formulas $\Gamma$ iff
every pertinent basic consequence of $\Gamma \cup \lbrace \alpha \rbrace$
is a basic consequence of $\Gamma$ (i.e. the addition of $\alpha$ to $\Gamma$ does not yield more pertinent formulas than with $\Gamma$ alone). This constitutes a pertinence consequence relation. More formally,

\begin{definition}
Let $\langle {\cal F}, {\cal V}, \models \rangle$ be a semantic structure
and $\mid\!\sim$ a relation on ${\cal P}({\cal F}) \times {\cal F}$.
\\
We say that $\mid\!\sim$ is a {\it pertinence consequence relation} (alias $X$-logic) iff
there exists ${\cal E} \subseteq {\cal F}$ such that $\forall \: \Gamma \subseteq {\cal F}$,
$\forall \: \alpha \in {\cal F}$,
$$
\Gamma \mid\!\sim \alpha \;\textrm{iff}\; \CC{\vdash}{\Gamma}{\alpha} \cap {\cal E} \subseteq \C{\vdash}{\Gamma}.
$$
In addition, if $\C{\vdash}{{\cal E}} = {\cal E}$, we say that $\mid\!\sim$ is {\it closed}.
\end{definition}
We introduce a new assumption about semantic structures
(in fact, simply a weak version of $(A3)$):

\begin{definition}
Suppose $\cal L$ is a language, $\vee$ a binary connective of $\cal L$, $\cal F$ the set of all wffs of $\cal L$,
and $\langle {\cal F}, {\cal V}, \models \rangle$ a semantic structure.
Then, define the following condition:
\begin{description}
\item[$(A4)$] $\forall \: \alpha, \beta \in {\cal F}$, $\M{\alpha \vee \beta} = \M{\alpha} \cup \M{\beta}$.
\end{description}
\end{definition}
We will show that when $(A4)$ is assumed,
then UC pivotal consequence relations are precisely closed pertinence consequence relations.
We need before Notation~\ref{PIVorTheo} and the very easy Proposition~\ref{PIVunion}
(which we will use implicitly in the sequel).

\begin{notation}\label{PIVorTheo}
Suppose $\cal L$ is a language, $\vee$ a binary connective of $\cal L$, $\cal F$ the set of all wffs of $\cal L$,
$\Gamma \subseteq {\cal F}$ and $\Delta \subseteq {\cal F}$. Then:
\\
$\Gamma \vee \Delta := \lbrace \alpha \vee \beta : \alpha \in \Gamma$ and $\beta \in \Delta \rbrace$.
\end{notation}

\begin{proposition} \label{PIVunion}
Suppose $\cal L$ is a language, $\vee$ a binary connective of $\cal L$, $\cal F$ the set of all wffs of $\cal L$,
$\langle {\cal F}, {\cal V}, \models \rangle$ a semantic structure satisfying $(A4)$,
$\Gamma \subseteq {\cal F}$, and $\Delta \subseteq {\cal F}$.
\\
Then, $\M{\Gamma} \cup \M{\Delta} = \M{\Gamma \vee \Delta}$.
\end{proposition}
\begin{proof}
Direction: ``$\subseteq$''.
\\
Suppose the contrary, i.e. suppose $\exists \: v \in \M{\Gamma} \cup \M{\Delta}$,
$v \not\in \M{\Gamma \vee \Delta}$.
\\
Then, $\exists \: \alpha \in \Gamma$, $\exists \: \beta \in \Delta$,
$v \not\in \M{\alpha \vee \beta}$.
\\
But, by $(A4)$, $v \in \M{\Gamma} \cup \M{\Delta} \subseteq \M{\alpha} \cup \M{\beta} = \M{\alpha \vee \beta}$, which is impossible.

Direction: ``$\supseteq$''.
\\
Suppose the contrary, i.e. suppose $\exists \: v \in \M{\Gamma \vee \Delta}$, $v \not\in \M{\Gamma} \cup \M{\Delta}$.
\\
Then, $\exists \: \alpha \in \Gamma$, $v \not\in \M{\alpha}$
and $\exists \: \beta \in \Delta$, $v \not\in \M{\beta}$.
\\
Therefore, by $(A4)$, $v \not\in \M{\alpha} \cup \M{\beta} = \M{\alpha \vee \beta}$.
\\
However $\alpha \vee \beta \in \Gamma \vee \Delta$.
Thus, $v \not\in \M{\Gamma \vee \Delta}$ which is impossible.\qed
\end{proof}

\begin{proposition} \label{PIVcodefisclosed}
Suppose $\cal L$ is a language, $\vee$ a binary connective of $\cal L$, $\cal F$ the set of all wffs of $\cal L$,
and $\langle {\cal F}, {\cal V}, \models \rangle$ a semantic structure satisfying $(A4)$.
\\
Then, UC pivotal consequence relations are precisely closed pertinence consequence relations.
\end{proposition}
\begin{proof}
Direction: ``$\subseteq$''.
\\
Let $\mid\!\sim$ be an UC pivotal consequence relation.
\\
Then, there is an UC SC choice function from $\bf D$ to ${\cal P}({\cal V})$ such that
$\forall \: \Gamma \subseteq {\cal F}$, $\C{\mid\!\sim}{\Gamma} = \T{\mu(\M{\Gamma})}$.
\\
Thus, by Proposition~\ref{PIVmupp}, there exists ${\cal I} \subseteq {\cal V}$ such that
${\cal V} \setminus {\cal I} \in {\bf D}$ and
$\forall \: \Gamma \subseteq {\cal F}$, $\C{\mid\!\sim}{\Gamma} = \T{\M{\Gamma} \cap {\cal I}}$.
\\
Define: ${\cal E} := \T{{\cal V} \setminus {\cal I}}$.
\\
Then, $\C{\vdash}{{\cal E}} = \T{\M{{\cal E}}} =
\T{\M{\T{{\cal V} \setminus {\cal I}}}} = \T{{\cal V} \setminus {\cal I}}
= {\cal E}$.
\\
In addition, as ${\cal V} \setminus {\cal I} \in {\bf D}$, we have $\M{{\cal E}} = \M{\T{{\cal V} \setminus {\cal I}}} = {\cal V} \setminus {\cal I}$.
\\
We show:
\\
$(0)$\quad $\forall \: \Gamma \subseteq {\cal F}$, $\forall \: \alpha \in {\cal F}$,
 $\Gamma \mid\!\sim \alpha$ iff $\CC{\vdash}{\Gamma}{\alpha} \cap {\cal E} \subseteq \C{\vdash}{\Gamma}$.
\\
Consequently, $\mid\!\sim$ is a closed pertinence consequence relation.

Direction: ``$\supseteq$''.
\\
Let $\mid\!\sim$ be a closed pertinence consequence relation.
\\
Then, there is ${\cal E} \subseteq {\cal F}$ such that ${\cal E} = \C{\vdash}{{\cal E}}$ and
$\forall \: \Gamma \subseteq {\cal F}$, $\forall \: \alpha \in {\cal F}$, $\Gamma \mid\!\sim \alpha$ iff
$\CC{\vdash}{\Gamma}{\alpha} \cap {\cal E} \subseteq \C{\vdash}{\Gamma}$.
\\
Define: ${\cal I} := {\cal V} \setminus \M{{\cal E}}$.
\\
Then, ${\cal V} \setminus {\cal I} = \M{{\cal E}} \in {\bf D}$.
\\
We will show:
\\
$(1)$\quad $\forall \: \Gamma \subseteq {\cal F}$, $\C{\mid\!\sim}{\Gamma} = \T{\M{\Gamma} \cap {\cal I}}$.
\\
Let $\mu$ be the choice function from $\bf D$ to ${\cal P}({\cal V})$ such that $\forall \: V \in {\bf D}$,
$\mu(V) = V \cap {\cal I}$.
\\
Then, $\forall \: \Gamma \subseteq {\cal F}$, $\C{\mid\!\sim}{\Gamma} = \T{\mu(\M{\Gamma})}$.
\\
In addition, by Proposition~\ref{PIVmupp}, $\mu$ is a UC SC choice function.
\\
Consequently, $\mid\!\sim$ is an UC pivotal consequence relation.
\\ \\
{\it Proof of $(0)$.}
Let $\Gamma \subseteq {\cal F}$ and $\alpha \in {\cal F}$.
Then:
\\$\Gamma \mid\!\sim \alpha$ iff
\\$\M{\Gamma} \cap {\cal I} \subseteq \M{\alpha}$ iff
\\$\M{\Gamma} \subseteq \M{\alpha} \cup ({\cal V} \setminus {\cal I})$ iff
\\$\M{\Gamma} \subseteq \M{\alpha} \cup \M{{\cal E}}$ iff
\\$\M{\Gamma} \subseteq \M{\Gamma \cup \lbrace \alpha \rbrace} \cup \M{{\cal E}}$ iff
\\$\M{\Gamma} \subseteq \M{(\Gamma \cup \lbrace \alpha \rbrace) \vee {\cal E}}$ iff
\\$\T{\M{(\Gamma \cup \lbrace \alpha \rbrace) \vee {\cal E}}} \subseteq \T{\M{\Gamma}}$ iff
\\$\T{\M{\Gamma \cup \lbrace \alpha \rbrace} \cup \M{{\cal E}}} \subseteq \T{\M{\Gamma}}$ iff
\\$\T{\M{\Gamma \cup \lbrace \alpha \rbrace}} \cap \T{\M{{\cal E}}} \subseteq \T{\M{\Gamma}}$ iff
\\$\CC{\vdash}{\Gamma}{\alpha} \cap \C{\vdash}{{\cal E}} \subseteq \C{\vdash}{\Gamma}$ iff
\\$\CC{\vdash}{\Gamma}{\alpha} \cap {\cal E} \subseteq \C{\vdash}{\Gamma}$.
\\ \\
{\it Proof of $(1)$.}
Let $\Gamma \subseteq {\cal F}$ and $\alpha \in {\cal F}$. Then:
\\
$\Gamma \mid\!\sim \alpha$ iff
\\$\CC{\vdash}{\Gamma}{\alpha} \cap {\cal E} \subseteq \C{\vdash}{\Gamma}$ iff
\\$\CC{\vdash}{\Gamma}{\alpha} \cap \C{\vdash}{{\cal E}} \subseteq \C{\vdash}{\Gamma}$ iff
\\$\T{\M{\Gamma \cup \lbrace \alpha \rbrace}} \cap \T{\M{{\cal E}}} \subseteq \T{\M{\Gamma}}$ iff
\\$\T{\M{\Gamma \cup \lbrace \alpha \rbrace} \cup \M{{\cal E}}} \subseteq \T{\M{\Gamma}}$ iff
\\$\T{\M{(\Gamma \cup \lbrace \alpha \rbrace) \vee {\cal E})}} \subseteq \T{\M{\Gamma}}$ iff
\\$\M{\Gamma} \subseteq \M{(\Gamma \cup \lbrace \alpha \rbrace) \vee {\cal E})}$ iff
\\$\M{\Gamma} \subseteq \M{\Gamma \cup \lbrace \alpha \rbrace} \cup \M{{\cal E}}$ iff
\\$\M{\Gamma} \subseteq \M{\alpha} \cup \M{{\cal E}}$ iff
\\$\M{\Gamma} \cap ({\cal V} \setminus \M{{\cal E}}) \subseteq \M{\alpha}$ iff
\\$\M{\Gamma} \cap {\cal I} \subseteq \M{\alpha}$.\qed
\end{proof}

%%%%%%%%%%%%%%%%%%%%%%%
\section{Conclusion}\label{PIVconclu}

We provided, in a general framework, characterizations for families
of pivotal(-discriminative) consequence relations.
We showed, in an infinite classical framework, that there is no normal characterization
for the family of all pivotal consequence
relations. And, we showed that UC pivotal consequence relations are precisely
those $X$-logics such that $X$ is closed under the basic entailment.
Beyond the contributions, an interest of the present paper is to give an example
of how the techniques developed in \cite{Bennaim1} (in particular in the discriminative case)
can be adapted to new properties (here Strong Coherence in the place of Coherence).
So naturally, we turn now to conclusions similar to those of \cite{Bennaim1}.
In many cases, our conditions are purely syntactic.
In fact, when the choice functions under consideration are not necessarily definability preserving,
we provided solutions with semi-syntactic conditions.
We managed to do so thanks to Lemmas~\ref{PIVmufestSCf} and \ref{PIVpivgen}.
An interesting thing is that we used them both in the plain and the discriminative versions.
This suggests that they can be used in yet other versions.
In addition, Lemmas~\ref{PIV3sim} and \ref{PIVPrf2ConArgSyn}
have been applied both here and previously in \cite{Bennaim1}
to characterize families of consequence relations defined in the discriminative manner by DP choice functions.
But, \cite{Bennaim1} is about coherent choice functions, whilst the present paper is about
strongly coherent choice functions. This suggests that these lemmas can be applied with yet
other properties.

%%%%%%%%%%%%%%%%%%%%%%%
\section{Acknowledgments}

I acknowledge Karl Schlechta, David Makinson, and Nicolas Ollinger for useful comments and suggestions.

\bibliography{generalBennaim}
\bibliographystyle{alpha}
\end{document}